\documentclass[acmsmall,nonacm]{acmart}

\renewcommand\footnotetextcopyrightpermission[1]{} 
\AtBeginDocument{%
  \providecommand\BibTeX{{%
    \normalfont B\kern-0.5em{\scshape i\kern-0.25em b}\kern-0.8em\TeX}}}

\acmJournal{JACM}
\acmVolume{37}
\acmNumber{4}
\acmArticle{111}
\acmMonth{8}

\usepackage{tabularx}
\usepackage{makecell}

\usepackage[inline]{enumitem}

\usepackage{multirow}

\usepackage{soul}
\usepackage{color}

\usepackage{xcolor,colortbl}

\usepackage{pifont}

\def\ccn{\cellcolor{red!25}{\ding{55}}}
\def\ccy{\cellcolor{green!25}{\ding{51}}}
\def\cco{\cellcolor{yellow!25}{\ding{109}}}

\usepackage[flushleft]{threeparttable}
\usepackage{longtable}

\usepackage{tikz}

\newcommand\bparagraph[1]{\vspace{1.0mm} \noindent\textbf{#1}}

\usepackage{fontawesome}
\newcommand{\unityIcon}{\includegraphics[scale=0.02]{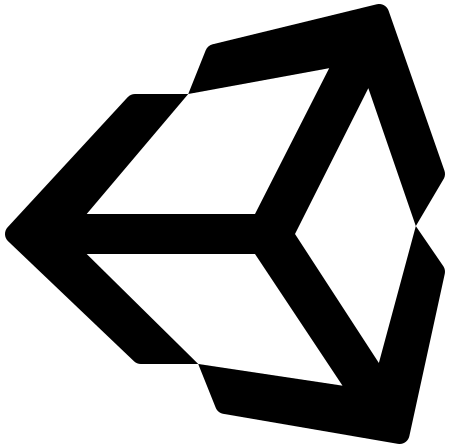}}%
\newcommand{\unityIconTiny}{\includegraphics[scale=0.015]{figures/icons8-unity-500.png}}%

\begin{document}

\title{Mobile Augmented Reality: User Interfaces, Frameworks, and Intelligence}

\author{Jacky Cao}
\affiliation{%
  \institution{University of Oulu}
  \city{Oulu}
  \country{Finland}}
\email{jacky.cao@oulu.fi}

\author{Kit-Yung Lam}
\affiliation{%
  \institution{The Hong Kong University of Science and Technology}
  \country{Hong Kong}}
\email{kylambd@connect.ust.hk}

\author{Lik-Hang Lee}
\affiliation{
  \institution{KAIST}
  \city{Daejeon}
  \country{Republic of Korea}}
\email{likhang.lee@kaist.ac.kr}

\author{Xiaoli Liu}
\affiliation{%
  \institution{University of Helsinki}
  \city{Helsinki}
  \country{Finland}}
\email{xiaoli.liu@helsinki.fi}

\author{Pan Hui}
\affiliation{%
  \institution{The Hong Kong University of Science and Technology}
  \country{Hong Kong;}
  \institution{University of Helsinki}
  \country{Finland}
}
\email{pan.hui@helsinki.fi}

\author{Xiang Su}
\affiliation{%
  \institution{Norwegian University of Science and Technology}
    \country{Norway;}
  \institution{University of Oulu}
  \country{Finland}}
\email{xiang.su@ntnu.no}

\begin{abstract}
Mobile Augmented Reality (MAR) integrates computer-generated virtual objects with physical environments for mobile devices. MAR systems enable users to interact with MAR devices, such as smartphones and head-worn wearables, and performs seamless transitions from the physical world to a mixed world with digital entities. These MAR systems support user experiences by using MAR devices to provide universal accessibility to digital contents. Over the past 20 years, a number of MAR systems have been developed, however, the studies and design of MAR frameworks have not yet been systematically reviewed from the perspective of user-centric design. This article presents the first effort of surveying existing MAR frameworks (count: 37) and further discusses the latest studies on MAR through a top-down approach: 
\begin{enumerate*}
    \item MAR applications;
    \item MAR visualisation techniques adaptive to user mobility and contexts;
    \item systematic evaluation of MAR frameworks including supported platforms and corresponding features such as tracking, feature extraction plus sensing capabilities;
    and \item underlying machine learning approaches supporting intelligent operations within MAR systems.
\end{enumerate*}
Finally, we summarise the development of emerging research fields, current state-of-the-art, and discuss the important open challenges and possible theoretical and technical directions. This survey aims to benefit both researchers and MAR system developers alike.
\end{abstract}

\ccsdesc[500]{Human-centered computing~Mixed / augmented reality}
\ccsdesc[300]{Software and its engineering~Development frameworks and environments}
\ccsdesc[300]{Computing methodologies~Machine learning approaches}
\ccsdesc[300]{Human-centered computing~Graphical user interfaces}

\keywords{mobile augmented reality, user interactions, development framework, artificial intelligence}

\maketitle

\section{Introduction} \label{sec:introduction}
Over the past several decades, Augmented Reality (AR) has evolved from interactive infrastructures in indoors or fixed locations to various mobile devices for ubiquitous access to digital entities~\cite{lee2018interaction}. We consider the first research prototype of MAR to be the \textit{touring machine system}~\cite{touring-machine-10.1145/151233.151239} that offers cues of road navigation on a school campus. After 20 years of development, MAR devices have shrunken from the size of huge backpacks to lightweight and head-worn devices. The interfaces on wearable head-worn computers have shifted from micro-interactions displaying swift (e.g., Google Glass) and small-volume contents to enriched holographic environments (e.g., Microsoft HoloLens)~\cite{lam2019m2a, lee2019hibey}. MAR offers interactive user experiences on mobile devices by overlaying digital contents, e.g., computer-generated texts, images, and audio, on top of physical environments. Numerous MAR frameworks have been developed to support the displaying of such enriched real-world environments by managing different sensors and components for tracking physical objects and various user interaction features (e.g., anchors and ray-casting), as well as offloading computational demanding tasks to remote servers. Nowadays, MAR, albeit primarily on smartphones, has been employed in various sectors, for instance, facilitating teaching and learning~\cite{sayed2011arsc, martin2010design, lin2013investigation, kamarainen2013ecomobile, lu2015integrating}, surgery training and reducing human errors during surgical operations~\cite{bernhardt2017status, kersten2013state, liao20103}, as well as virtual tour guides during sight-seeing~\cite{mobilemarkettech:tuscany, perey:basel, han2013dublin, wei2014haptic}. The multitude of MAR applications has proven its practicality in becoming a ubiquitous interface type in the real world. With the recent advancements of Artificial Intelligence (AI) and Internet of Things (IoT) devices, numerous mobile equipment and IoT devices construct smart and responsive environments. Users leverage head-worn computers to access a multitude of intelligent services through MAR. Accordingly, we advocate that MAR has to develop adaptability, or \textit{adaptive MAR frameworks}, to address the on-demand user interactions with various IoT devices in the aforementioned smart environments, such as drones, robotics, smart homes, and sensor networks.  
 
One critical research challenge to realise the MAR vision is the provisioning of seamless interactions and supporting end-users to access various MAR services conveniently, for instance, through wearable head-worn computers and head-mounted displays (HMDs).HMDs are technological enablers for user interactions within augmented environments. That is, these headsets project digital contents in the form of windows, icons, or more complex 3D objects (e.g., virtual agents/avatars) on top of views of the physical world. Users with such wearable head-worn computers can access various AR services ubiquitously and new forms of services and user experiences will spread and penetrate into our daily routines, such as purchasing tickets from transport service providers, setting privacy preferences and control on small-size smart cameras and speakers, and buying snacks and drinks from convenience stores. The interfaces of MAR experiences are regarded as ubiquitous service points to end users.

AI is crucial to support features of high awareness to user physical surroundings and user context to support MAR experiences, and in order to achieve user-oriented services. Considerations of user performance, user acceptance, and qualitative feedback in such MAR systems make adaptive MAR user interfaces (UIs) a necessity in rich yet complex physical environments. To support the creation of MAR experiences, several frameworks are available where developers can deploy pre-built systems, which already contain features such as displaying augmented objects, performing environment analysis, and supporting collaborative AR with multiple users. From environment analysis found in MAR frameworks to adaptive MAR UIs, and to enable low-latency and good user experiences, AI methods are generally required to execute these tasks. Nowadays, more commonly, machine learning (ML) methods are typically used due to their ability to analyse and extract several layers of information from data (e.g., geographical data, user data, image frames taken from physical environments) in an efficient manner. 

Thus, this survey examines the recent works of MAR with focusing on developments and synthesis of MAR applications, visualisations, interfaces, frameworks, and applied ML for MAR. We also strive to move beyond the individual applications of MAR and seek research efforts towards highly intelligent and user-oriented MAR frameworks that potentially connects with not only digital and virtual entities, but also physical daily objects. We outline key challenges and major research opportunities for user-oriented MAR UIs and the AI methods for supporting such UIs. The contributions of this survey paper are as follows:
\begin{enumerate}
    \item Provisioning of a comprehensive review of the latest AR applications and framework-supported interface designs in AR in both hand-held and head-worn AR system scenarios;
    \item Provisioning of a comprehensive review of existing AR frameworks and SDKs, which explores the supported hardware platforms and features of the MAR frameworks; 
    \item Provisioning of a comprehensive review of relevant ML methods to be applied in key AR components of MAR frameworks;
    \item A research agenda for the development of future adaptive AR user interactions concerning systematic MAR frameworks and ML-assisted pipelines.
\end{enumerate}                             

Among our main calls to action in the agenda are issues related to AR experiences, such as investigating the feasibility of AR interfaces in user-orientated and high adaptive manners, developing ML methods for MAR experiences in dynamic and complicated environments, and advancing MAR systems for two-way and seamless interactions with all intelligent objects in evolving MAR landscapes. 

\subsection{Overview of Mobile Augmented Reality} \label{sec:overview}
\begin{figure}
    \centering
    \includegraphics[width=0.88\textwidth]{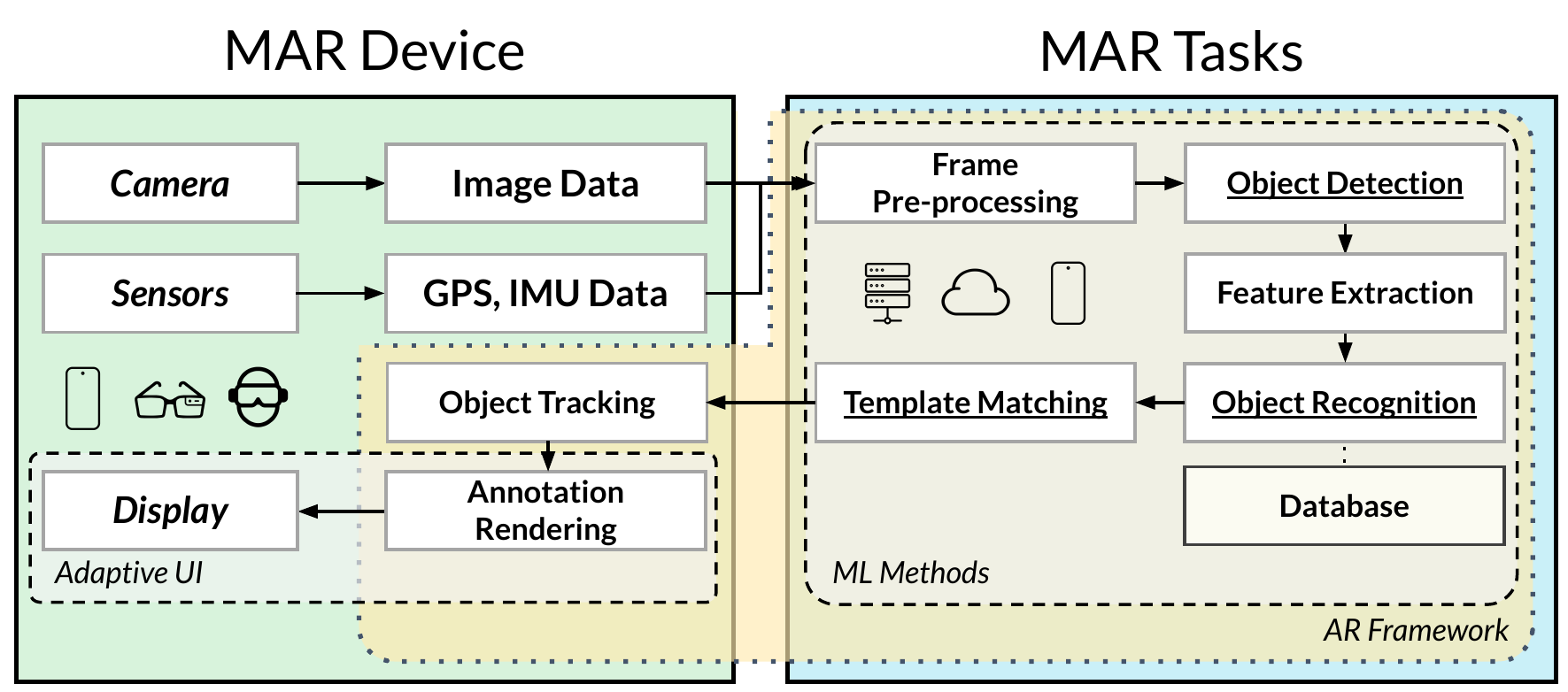}
    \caption{A typical MAR system pipeline, splitting into MAR device and MAR tasks. ML methods can be utilised in the majority of required MAR tasks.}
    \label{fig:mar_pipeline}
\end{figure}

Figure \ref{fig:mar_pipeline} presents the typical pipeline of an MAR system. The components in MAR systems can be approximately grouped into \textit{MAR Device} and \textit{MAR Tasks}, where the tasks may be executed on device, or on an external server. Generally, hardware sensors (i.e., cameras, GPS and IMU sensors, and LiDAR modules) on the MAR device first captures images and sensor data which are then fed into the MAR tasks. The first of which is frame pre-processing, where data is cleaned and prepared to be inputted into the analysis tasks to then retrieve information on which objects are in the environment, e.g., object detection, feature extraction, object recognition, and template matching tasks. These tasks can be classed as the \textit{ML Tasks}, where ML techniques are used to complete the tasks in a more efficient manner due to their complexity. After data analysis, successful results and associated annotations are returned to the MAR device for object tracking and annotation rendering on the client. Together with ML methods, this can form an overall \textit{AR Framework} where developers can employ pre-built frameworks to create their own MAR applications. The annotation rendering and display components of the MAR device can be grouped together as part of \textit{Adaptive UI}, the contents and interaction with the rendered augmentations are important for users as this allows them to experience MAR. Therefore, the scope of this survey primarily aims to address the following questions, which are missing in the existing surveys (detailed in Section~\ref{sec:related}): 
\begin{enumerate*}
    \item What features and functions are available in MAR frameworks nowadays?
    \item What are the latest efforts of AR research and accordingly, the prominent gap between these efforts and the identified frameworks?
    \item Then based on the above questions, how MAR engineers and practitioners design the next generation of MAR frameworks that bridge the user requirements?
\end{enumerate*}

\subsection{Related Surveys and Selection of Articles}\label{sec:related} 
This subsection briefly outlines previous relevant surveys in various domains, i.e. user interfaces, frameworks, and intelligence. One of the earliest AR surveys, focusing on AR systems, applications, and supporting technologies, is proposed by Carmigniani et al. \cite{carmigniani2011augmented}. Later on, Lee and Hui~\cite{lee2018interaction} focus on the user interaction techniques on MAR devices (i.e., smartglasses). A recent survey by Chatzopoulos \textit{et al.} \cite{chatzopoulos2017mobile} present a comprehensive state-of-the-art survey of MAR application fields, how MAR user interfaces can be developed, what key components MAR systems consists of, and how data is managed in MAR systems. Zollmann \textit{et al.} \cite{zollmann2020visualization} re-organise the recent works of different components in the visualisation pipelines of AR systems and presents the findings in three main areas, including filtering, mapping, and rendering. Marneanu \textit{et al.} \cite{iJIM3974} provide an early evaluation of AR frameworks for Android developments. Recent AR development platforms on both Android and iOS platforms have been generally discussed in some development communities, but have never been evaluated in a systematic and scientific way. Herpich \textit{et al.} \cite{herpich2017comparative} perform a comparative analysis of AR frameworks, albeit for the scenario of developing educational applications. Finally, utilisation of ML in different core components of MAR applications are discussed in several surveys. For example, Krebs \textit{et al.} \cite{krebs2017survey} present a survey of leveraging deep neural networks for object tracking. Liu \textit{et al.} \cite{liu2020deep} focuse on deep learning for object detection. Yilmaz \textit{et al.} \cite{yilmaz2006object} survey object tracking. Sahu \textit{et al.} \cite{sahu2020artificial} conduct a comprehensive study on fundamental ML techniques for computational pipelines of AR systems (e.g., camera calibration, detection, tracking, camera pose estimation, rendering, virtual object creation, and registration), in AR-assisted manufacturing scenarios. Our article addresses the gaps in these existing surveys and provides a comprehensive survey on MAR interface design, frameworks and SDKs, and relevant ML methods utilised in key AR components.

The articles are initially selected by searching recent research articles using a small number of keyword seeds on Google Scholar, IEEE Xplore, ACM Digital Library, Scopus, Web of knowledge, etc. The queries combine keywords from the set \textit{MAR, ubiquitous interaction, user interaction, AR adaptive UI, AR visualisation techniques, AR camera calibration, camera pose estimation, AR remote collaboration, AR user collaboration, MAR frameworks, object detection, object tracking,  object recognition,  rendering, etc}. 
The initial article set was expanded by taking into consideration articles that cite or are cited by the articles within this set. We further complement the set of articles by adding prominent research featured in media to cover a comprehensive set of important publications in this area. This process was continued until no new articles were found. We then discussed the papers among the authors and analysed the most relevant and important of the selected articles by reading the abstract and the main findings of the papers. Papers that were less relevant for the scope of the survey were filtered out during this process. The selected papers form the core of this survey and we have performed continuous updates during the survey writing process to cover papers published since the start of our process.

\subsection{Structure of the survey}
The remainder of this article is organised as follows. Section \ref{sec:applications} summarises significant application areas which MAR can fulfil. We present detailed discussion and analysis on the development and recent advances of adaptive UI for MAR, MAR frameworks, and ML for MAR, in Section \ref{sec:adaptiveui}, Section \ref{sec:framework}, and Section \ref{sec:mlmethods}, respectively. Finally, we discuss research challenges and future directions in Section \ref{sec:future_research} and conclude the paper in Section \ref{sec:conclusions}. 
\section{MAR Applications}\label{sec:applications} 
Over the past few years, AR has gained wide spread media attention and recognition due to easy access through modern smartphone applications. As an emerging technology, research and development of MAR have been seen in several different fields. The versatility and ease-to-use of MAR highlight the potential for the technology to become a ubiquitous part of everyday life. In 2017 alone, the revenue of the AR market was worth \$3.5 billion \cite{statista:ar_market} and by 2023 this value could reach up to \$70--\$75 billion \cite{digicapital:ar2}. A study conducted by SuperData and Accenture Interactive found that 77\% of AR users are likely to use AR as a content viewer (e.g., for file viewing), 75\% are likely to use AR in shopping scenarios, and 67\% are likely to use AR for gaming entertainment. It reveals the importance of considering AR user interfaces, the underlying AR system, and how users can continually be encouraged to use AR. Among the different AR scenarios, the requirements of user interface are different between each of scenarios. In this section, we study several particular AR scenarios and applications with an emphasis on the types of user interfaces employed. This leads to a greater understanding as to why AR has been such a prominent technology and why it continues to attract research attention.

\subsection{Entertainment \& Tourism}
AR technologies push the boundaries of entertainment and tourism. Pok\'{e}mon GO~\cite{rauschnabel2017adoption} is a mobile game which uses AR to bring fictional Pok\'{e}mon characters into the real world. Released in July 2016, at the height of the game's popularity, there were 45 million users and generated a first month of revenue of \$207 million \cite{businessofapps:pokemongo}. This has led to the creation of similar concept AR gaming apps, such as Harry Potter: Wizards Unite \cite{harrypotter:ar_game}. AR gaming is not restricted to enhancing optical camera feeds. Audio AR experiences is another type of augmentation, for example, as found in the running app, Zombies, Run!~\cite{zombiesrun}. Social media applications, such as Snapchat and Instagram allow users to apply ``face filters'' or ``lenses'' to themselves using AR \cite{snapchat:lenses}. In addition, Snapchat and social media application TikTok, have utilised the LiDAR sensor in the 2020 generation of iPhones to create environmental AR effects \cite{gartenberg_tiktok_2021}. 

AR could act as digital tour guides for landmarks or cultural sites in tourism applications. Several projects create interesting MAR applications, which are offered to visitors to Tuscany, Italy \cite{mobilemarkettech:tuscany} and Basel, Switzerland \cite{perey:basel} with city navigation guidance, sight-seeing destinations, and general information. The design of such applications is traveller-centric so information is tailored to that specific audience, for example, providing multi-language functionality \cite{han2013dublin}, offering alternative haptic or audio AR experiences \cite{wei2014haptic}, or supplying information which is suited to an individual's knowledge level \cite{kounavis2012enhancing}. Wide-spread adoption of AR has yet to emerge. However, with continual exposure and improving technologies, there is no doubt that AR will become a prevalent part of everyday life in several entertainment and tourism sectors. 

\subsection{Education}
Over the past 30 years, classrooms have been evolving through the introduction of various technologies. From computers in classrooms \cite{dwyer1991changes} to lecture theatres with projection systems \cite{nantz1998lecturing}, classroom-based learning is now incorporating AR technology as an educational tool. Educational AR can be categorised as subject specific AR applications and separate AR teaching tools. Subject specific AR applications are developed for promoting a specific set of knowledge, and the user interfaces for these applications are well designed for presenting interactive contents. AR teaching tools apply AR technologies for improving the efficiency of teaching. The user interface for AR teaching tools is designed for facilitating the communications between users during education activities. Advantages are shown to arise from the use of AR in educational settings, for example, the individual skills of students can be improved, such as an increased ability to understand and handle information \cite{sayed2011arsc, kamarainen2013ecomobile}, enhanced confidence \cite{lu2015integrating}, and a development in spatial abilities \cite{kaufmann2002mathematics, martin2010design, lin2013investigation}. 

AR could enhance student-teacher communications within classrooms. Zarraonandia \textit{et al.~}\cite{zarraonandia2013augmented} suggest providing teachers with HMDs and students with devices which can alert the teacher if a student is struggling. Holstein \textit{et al.~}\cite{holstein2019designing} design intelligent tutoring systems as a rich formative assessment tool, which can augment teachers’ perceptions of student learning and behaviour in real time. For students with disabilities and learning difficulties, the teaching can be tailored to individual specific needs. For example, pupils with attention deficit hyperactive disorder could use AR to help emphasise and improve engagement, motivation, and content visualisation \cite{ab2012providing}. Martín-Gutiérrez \textit{et al.~}\cite{martin2015augmented} present a collaborative and autonomous 3D AR learning experiment, which combines every learning process from an electrical machines course in an electrical engineering degree. Their AR tool allows interactive and autonomous studying as well as collaborative laboratory experiments with other students and without assistance from teachers. 

\subsection{Healthcare}
AR applications in the medical field offer new approaches for patient-doctor relationships, treatments, and medical education. For example, AR is categorised as an assisting technology to support individuals who have Alzheimer's disease to help identify objects and people, provide reminders to take the correct medication, and aid caregivers in locating patients \cite{kanno2018augmented}. Alternatively, current medical imaging modalities, such as ultrasound and CAT scans, can be greatly enhanced through scan virtualisation into 3D models and their superposition onto physical body parts \cite{peters2018mixed}. With a scenario such as a pregnant mother receiving an ultrasound, the scan could be casted in real-time to several different observers wearing HMDs within the same room. All participants could receive the same ultrasound hologram on top of the patient, but midwives and other medical staff could be viewing additional information, such as past medical history and other vital signs \cite{chen1994case, mahmood2018augmented}. Novarad OpenSight presents one particular Microsoft HoloLens application product for CT images~\cite{gibby2019head}, allowing a user to view virtual trajectory guides and CT images which are superimposed on physical objects as two or three dimensional augmentations. 

One of the most significant areas of AR for medical research pertains to utilising AR for surgery in various different stages of treatment. Firstly, a surgery can be explored and planned before execution, which allows surgeons to develop the most optimal treatment strategy \cite{bornik2003computer}. Once the surgery procedure begins, image guidance can be used to provide surgeons with navigation information so their attention is not drawn away from the operative field \cite{fuchs1998augmented, drouin20187}. In addition, mixed reality image-guided surgery (IGS) has been frequently used for surgery training \cite{bernhardt2017status, kersten2013state} and minimally invasive surgeries. In IGS, surgical instruments are tracked and visualised with respect to patient-specific data sets to guide surgeons while in the operating theatre. Several surgical fields have already begun to apply AR, for example, neurosurgery, general surgery, and orthopedic surgery \cite{shuhaiber2004augmented}. 

Adaptive user interface techniques are critical for surgery scenarios. To support this, several techniques, including label placement, occlusion representation, and registration error adaptation, are extensively studied. For example, Fuchs \textit{et al.~}\cite{fuchs1998augmented} use depth cues of occlusion to aid in the perception of the order of objects. Specifically, a pixel is painted only if it lies above the surface of the closest object. In the AR system of Wieczorek \textit{et al.~}\cite{wieczorek2010gpu}, occlusion handling of the surgeon's hand gives a more realistic perception and blending of realities. Pratt \textit{et al.~}\cite{pratt2018through} demonstrate that AR can assist accurate identification, dissection, and execution of vascular pedunculated flaps during reconstructive surgery.
\section{Composing AR visualisation and user environments} \label{sec:adaptiveui}
Among all the AR applications (Section~\ref{sec:applications}), the general AR experience is to display virtual entities overlaid on top of the physical world. For producing immersive AR experiences, information management and rendering are key processes for delivering virtual entities. Collaborative AR system enables users to collaborate with each other within the physical and virtual environments. Numerous MAR frameworks are designated for the aforementioned purposes. Before we describe the existing MAR frameworks (Section~\ref{sec:framework}), this section provides an overview of the relevant concepts and existing works in UI adaptability as well as device and user collaborations. We also explain the existing metrics of measuring the effectiveness of AR. 

\subsection{Adaptive UI for MAR}
After data is prepared for visualisation, there are three main steps in the visualisation pipeline \cite{zollmann2020visualization}, including
\begin{enumerate*}
    \item filtering,
    \item mapping,
    and \item rendering.
\end{enumerate*}
Zollmann \textit{et al.~}\cite{zollmann2020visualization} perform a comprehensive review organised by this visualisation pipeline. We adopt their approach for classifying the characteristics of visualisation techniques and we further investigate the context-aware adaptive interface techniques which adapt to a person, task, or context environment. Several algorithms, such as ML-based algorithms, have been applied in adaptive user interfaces for improving human-interface interactions. They are designed to aid the user in reaching their goals efficiently, more easily, or to a higher level of satisfaction. We evaluate these works by considering the user experience measurement perspectives, and summarise four adaptive user interface techniques~\cite{julier2003adaptive, zollmann2020visualization} commonly found in the existing frameworks, with the following requirements for the implementations of each set of interfaces:

\textbf{Information Filtering and Clustering (InfoF):} Overload and clutter occur when a large amount of information is rendered to users. Complex visualisation of augmented contents negatively affects visual search and other measures of visual performance. In order to produce better user interfaces, there are two approaches to reduce the complexity of interfaces by decreasing the amount of displayed contents. The first approach is information filtering which can be implemented by developing a filtering algorithm that utilises two steps, namely, a culling step followed by a detailed refinement step. Another approach is information clustering \cite{tanaka2008information}, which groups together and represents the information by their classification.

\textbf{Occlusion Representation and Depth Cues (OcclR):} Occlusion representation provides depth cues to determine the ordering of objects. Users can identify the 3D locations of other physical objects when large structures occlude over or under each other. Julier \textit{et al.~}\cite{julier2000information} suggest three basic requirements for depth cues, including
\begin{enumerate*}
    \item the ability to identify and classify occluding contours;
    \item the ability to calculate the level of occlusion of target objects, and parameterising the occlusion levels if different parts of the object require this;
    and \item the ability to use perceptually identified encodings to draw objects at different levels of occlusion.
\end{enumerate*} 

\textbf{Illumination Estimation (IEsti):} Light estimation is another source of information depicting the spatial relations in depth perception, which enhances visual coherence in AR applications by providing accurate and temporally coherent estimates of real illumination. The common methods for the illumination estimation can be classified into auxiliary information or non-auxiliary information.

\textbf{Registration Error Adaptation (RegEA):} Another mapping step of the rendering pipeline. Trackers are imprecise and time-varying registration often exists. Therefore, the correct calibration of devices and displays is difficult. This subsequently leads to graphical contents not always aligning perfectly with their physical counterparts. Accordingly, the UI should be able to dynamically adapt while visualising the information. Ambiguity arises when the virtual content is not clearly interacting with the context environment around users.

\textbf{Adaptive Content Placement (AdaCP):} AR annotations are dynamic 3D objects, which can be rendered on top of physical environments. Labels have to be drawn with respect to the visible part of each object, in order to avoid confusing and ambiguous interactions.

\bparagraph{InfoF}:
For some context-aware AR applications, information clutter is prevalent in outdoor environments due to the complexity of them. Exploring and searching for information on AR screens become indispensable tasks for users. Information filtering is a necessary technique in modern AR systems~\cite{MarkusTatzgern2016}, especially in large and complicated environments, where information overload is significant. Without intelligent automation of filtering and selection tools, the display would always lead to difficulty in reading information \cite{khan2015rise}. There are three main methods for filtering information, including
\begin{enumerate*}   
    \item spatial filters, 
    \item knowledge-based filters,   
    and \item location-based filters~\cite{MarkusTatzgern2016}.
\end{enumerate*}
Spatial filters select information displayed on screens or in the object space based on physical dimension rules. These filters require user interactions to investigate the entirety of virtual contents, for example, users have to move their MAR devices to view large 3D models. But the method only works locally in a small region. The immersive AR application always applies the spatial filters to exclude the information that is out of the user's view frustum. Knowledge-based filters enable user preferences to be the filtering criteria \cite{MarkusTatzgern2016}. Expert knowledge filters embedded behaviour and knowledge in the coding, regulating a series of data structures used by the system to infer recommendations and output the items satisfying user requirements. Such knowledge coding can be done in different ways, such as rules in a rule-based systems~\cite{plaza2014arlodge}. Finally, location-based filters are a hybrid method combining spatial and knowledge-based filters. As new sensors are embedded in modern AR headsets, such as gaze sensors, the user bio-information and context information can be used for filtering \cite{ajanki2011augmented}.

\bparagraph{OcclR}:
Comprehensive AR systems not only track sparse geometric features but also compute depth maps for all pixels when visualising occluded objects or floating objects in AR~\cite{julier2003adaptive}. Depth maps provide depth values for each pixel in captured scenes, they are essential for generating depth cues, and they help users to understand their environment and to aid interactions with occluded or hidden objects. Recent AR frameworks, such as Google ARCore~\cite{google:arcore} and Apple ARKit~\cite{apple:ar}, provide depth map data for enabling depth cue features in MAR applications \cite{du2020depthlab}. Physical and virtual cues are two options for producing depth cues in AR applications to support depth perception \cite{zollmann2020visualization}. Physical cues can be used to rebuild natural pictorial depth cues~\cite{wither2005pictorial}, such as occlusion or shadows. Integrating depth maps and RGB camera images together can provide the necessary natural pictorial depth cues \cite{zollmann2020visualization}. Subsequently, virtual cues and graphical aids are generated by applications to provide similar depth cues as physical alternatives. ``X-ray vision'' is a technique for virtual cues frequently used for perceiving graphics as being located behind opaque surfaces. DepthLab \cite{du2020depthlab} is an application using ARCore's Depth API, enabling both physical and virtual cues, helping application developers to effortlessly integrate depth into their AR experiences. DepthLab implements the depth map and depth cues for at least six kind of interactions in AR, including
\begin{enumerate*}
    \item oriented reticles and splats,
    \item ray-marching-based scene relighting,
    \item depth visualisation and particles,
    \item geometry-aware collisions,
    \item 3D-anchored focus and aperture effect,
    and \item occlusion and path planning \cite{du2020depthlab}. 
\end{enumerate*}
Illumination estimation is typically achieved with two traditional approaches, including:
\begin{enumerate*}
    \item Methods utilising auxiliary information which leverages RGB-D data or information acquired from light probes and the methods can be an active method like the fisheye camera used by K\'{a}n \textit{et al.}~\cite{kan2013differential} or a passive method like reflective spheres used by Debevec~\cite{10.1145/1401132.1401175}, and 
    \item Estimating the illumination using an image from the main AR camera without the need of having an arbitrary known object in the scene. The auxiliary information can also be assumptions of some image features which are known to be directly affected by illumination, or simpler models like Lambertian illumination. Shadows, gradient of image brightness~\cite{boom2013point, boom2017interactive, kasper2017light} and shading~\cite{gruber2012real, gruber2015image,jiddi2016reflectance, lopez2013multiple, richter2016instant, whelan2016elasticfusion} are the typical image features used for estimating illumination direction. 
\end{enumerate*}

\bparagraph{RegEA}:
Addressing registration and sensing errors is a fundamental problem in building effective AR systems \cite{azuma1997survey}. Serious errors result in visual-kinesthetic and visual-proprioceptive conflicts. Such conflicts between different human senses may be a source of motion sickness. Therefore, the user interface must automatically adapt to changing registration errors.   
MacIntyre B. \textit{et al.~}\cite{macintyre2000adapting} suggest using Level Of-Error (LOE) object filtering for different representations of an augmentation to be automatically used as registration error changes. This approach requires the identification of a target object and a set of confusers \cite{macintyre2002estimating}. 
Afterwards, their approach calculates the registration errors for the target and all confusers. The delegation error convex hulls are used to bound the geometry of the objects. The hulls are constructed for two disjointed objects in the presence of substantial yaw error. The hull surrounding each object together with a suitable label is sufficient to direct the user to the correct object edges.

Registration error adaptation is a critical issue in safety-critical AR systems, such as for surgical or military applications. Recent AR frameworks provide real-time registration error adaption with precise IMU tracking data and camera image fusion algorithms, which minimises the registration error. Robertson and MacIntyre~\cite{robertson2004adapting} describes a set of AR visualisation techniques for augmentations that can adapt to changing registration errors. The first and traditional technique is providing general visual context of an augmentation in the physical world, helping users to realise the intended target of an augmentation. It can be achieved by highlighting features of the parent object, and showing more feature details as the registration error estimate increases. 
The second technique is presenting the detailed visual relationships between an augmentation and nearby objects in the physical world, by highlighting a unique collection of objects near the target of the augmentation in the physical world, and allowing the user to differentiate between the augmentation target and other parts of the physical world that are similar.

\bparagraph{AdaCP}:
Major label placement solutions include greedy algorithms, cluster-based methods, and screen subdivision methods \cite{azuma2003evaluating}. Other methods include making the links between objects and their annotations more intuitive, alleviating the depth ambiguity problem, and maintaining depth separation \cite{mcnamara2019information}. Labels have to be drawn with respect to each visible part of the object. Otherwise, the results are confusing, ambiguous, or even wrong. The implementation is simple computed axially-aligned approximations of the object projections, the visibility is then determined with simple depth ordering algorithms \cite{bell2001view}. Several research works focus on providing appropriate moving label dynamics to ensure that the temporal behaviour of the moving labels facilitates legibility. From these works, certain requirements arise, such as the ability to determine visible objects, the parameterisation of free and open spaces in the view plane to determine where and how content should be placed, and labels should be animated in real-time because the drawing characteristics should be updated on a per-frame basis~\cite{julier2003adaptive}. The requirements for adaptive content placement solutions are targeting to produce better user experiences.  

View management algorithms address label placement \cite{azuma2003evaluating}. Wither \textit{et al.~}\cite{wither2006using, wither2008fast} provide an in-depth taxonomy of annotations, especially regarding the location and permanence of annotations. Tatzgern \textit{et al.~}\cite{MarkusTatzgern2016} propose a cluster hierarchy-based view management system. Labels are clustered to create a hierarchical representation of the data and visualised based on the user’s 3D viewpoint. Their label placement employs the ``hedgehog labeling'' technique, which places annotations in real world space to achieve stable layouts. McNamara \textit{et al.~}\cite{mcnamara2016mobile} illustrate an egocentric view-based management system that arranges and displays AR content based on user attention. Their solution uses a combination of screen position and eye-movement tracking to ensure that label placement does not become distracting. 

For a comprehensive MAR system, current interfaces do not consider walking scenarios. Lages \textit{et al.~} \cite{lages2019walking} explore different information layout adaptation strategies in immersive AR environments. A desirable property of adaptation-based interface techniques is developed in their study. Adaptive content management is implemented in an MAR system, where the behaviours function in a modular system can combine individual behaviours, and a final minimal set of useful behaviours is proposed that can be easily controlled by the user in a variety of mobile and stationary tasks \cite{rakholia2018place,pei2019wa, jia2020semantic, hegde2020smartoverlays}.
 
\subsection{Collaborative UIs in Multi-user and Multi-device AR}
Adaptive AR UIs can serve as ubiquitous displays of virtual objects that can be shown anywhere in our physical surroundings. That is, virtual objects can be floating in the air on any physical background, which can be reached out to or manipulated by multiple users with their egocentric views~\cite{p1-10.5555/946248.946788}. Users engaged in their AR-mediated physical surroundings are encouraged to accomplish tasks in co-facilitated environments with shared and collaborative AR experiences amongst multiple users~\cite{p2-10.1145/3281505.3283392}. Multiple dimensions of AR collaborative UIs are discussed throughout various applications, for instance, working~\cite{p2-10.1145/3281505.3283392,p14-10.1145/3229089} and playful~\cite{p1-10.5555/946248.946788, p6-10.1145/3089269.3089281} contents, local/co-located~\cite{p4-10.1145/2839462.2856521} and remote~\cite{p3-10.1145/3355355.3361892} users, sole AR~\cite{p2-10.1145/3281505.3283392, p4-10.1145/2839462.2856521} and a mixture of AR and VR~\cite{p5-10.1145/3132818.3132822,p6-10.1145/3089269.3089281}, co-creation/co-editing by multiple users (i.e., multiple users at the front of AR scenes)~\cite{p2-10.1145/3281505.3283392, p4-10.1145/2839462.2856521}, supported and guided instruction (i.e., multiple users connecting to one user at the front of AR scenes)~\cite{p6-10.1145/3089269.3089281, p10-10.1145/2875194.2875204}, human-to-human interactions~\cite{p1-10.5555/946248.946788, p2-10.1145/3281505.3283392, p4-10.1145/2839462.2856521}, and the interaction between human and AR bots, such as tangible robots~\cite{p8-10.1145/3240508.3241390} and digital agent representatives~\cite{p9-10.1145/2522848.2522855}.

Multi-user collaborative environments has been a research topic in the domain of human-computer interaction, which has evolved from sedentary desktop computers~\cite{p11-10.1145/1188816.1188821} to mobile devices and head-worn computers~\cite{p2-10.1145/3281505.3283392, p3-10.1145/3355355.3361892, p6-10.1145/3089269.3089281, p14-10.1145/3229089}. The success of such collaborative environments needs to cope with several design challenges, including
\begin{enumerate*}
    \item high awareness of others' actions and intentions,
    \item high control over the interface,
    \item high availability of background information,
    \item information transparency amongst collaborators,
    and \item information transparency whilst not affecting other users \cite{p12-10.1145/2147783.2147784, p13-10.1145/1873561.1873568}.
\end{enumerate*}
These design challenges of awareness, control, availability, transparency, and privacy serve as fundamental issues enabling multiple users to smoothly interact with others in collaborative and shared environments. When the collaborative and shared environments are deployed to AR, features such as enriched reality-based interaction and high levels of adaptability are introduced. The additional design challenges extend from resolving multi-user experiences, as discussed previously, to reality-based interaction across various AR devices including AR/VR headsets, smartwatches, smartphones, tablets, large-screen displays, and projectors. These challenges place a priority on exploring the management of multiple devices and their platform restrictions, unifying the device-specific sensing and their interaction modalities, connecting the collaborative AR environments with physical coordinate systems in shared views~\cite{p14-10.1145/3229089}. The complexity of managing diverse devices leads to the need of an AR framework to systematically and automatically enable user collaborations in co-aligned AR UIs. It is important to note that the majority of evaluation frameworks focus on small numbers of quantitative metrics, such as completion time and error rate in an example of visual communication cues between an on-site operator and a remote-supporting expert ~\cite{communication-cues-10.1145/3290605.3300403}, and they neglect the multi-user responses to the physical environments.

\subsection{Metric Evaluations for AR UIs}
AR interfaces were originally considered for industrial applications, and the goodness or effectiveness of such augmentations were limited to work-oriented metrics~\cite{m1-10.1145/2037373.2037504}. A very early example refers to augmented information, such as working instructions for front-line labourers on assembly lines, where productivity, work quality, and work consistency are regarded as evaluation metrics~\cite{m8-10.1145/2836041.2836067}. However, these task-orientated metrics are not equivalent to user experience (e.g., the easiness of handling the augmentation), and neglects the critical aspects especially user-centric metrics. The multitudinous user-centric metrics are generally inherited from traditional UX design issues, which can be categorised into four aspects: 
\begin{enumerate*}
    \item user perception of information (e.g., whether the information is comprehensible, understandable, or can be easily learnt),
    \item manipulability (i.e., usability, operability), 
    \item task-oriented outcome (e.g., efficiency, effectiveness, task success),
    and \item other subjective metrics (e.g., attractiveness, engagement, satisfaction in use, social presence, user control) \cite{m4-ARIFIN2018648}.
\end{enumerate*}
When augmentations are displayed on handheld devices, such as smartphones and tablets, ergonomic issues, such as ease in AR content manipulation with two-handed and one-handed operations are further considered~\cite{m2-10.1145/2671015.2671019}. Nowadays, AR is deployed in real-world scenarios, primarily acting as marketing tools, and hence business-orientated metrics are further examined, such as utility, aesthetic, enjoyment, and brand perception~\cite{m6-10.1145/3385378.3385383, m15-APUX-10.1145/2846439.2846450}. Additionally, multi-user collaborative environments require remote connections in AR ~\cite{edge-10.1145/3304112.3325612}, and lately, the quality of experience (QoE) through computation offloading mechanisms to cloud or edge servers encounters new design challenges of AR UIs ~\cite{m10-10.1145/3344341.3368816, m11-9103475}. 

The user perception of AR information which leads to the comprehensibility of AR cues and the learnability of AR operations in reality-based interactions, has been further investigated as a problem of multi-modal cues, such as audio, video, and haptics in various enriched situations driven by AR~\cite{m3-10.1145/3385378.3385384}. The intelligent selection of AR information and adaptive management of information display and multi-modal cues are crucial to users' perception to AR environments~\cite{lam2019m2a, chatzopoulos2016readme}. This can be considered as a fundamental issue of interface plasticity. In the mixed contents between the digital and physical realities~\cite{m4-ARIFIN2018648}, the plasticity of AR interfaces refers to the compatibility of the information to physical surroundings as well as situations of users (i.e., context-awareness)~\cite{m5-10.1145/2617841.2620695,m16-cognition-10.1145/3332165.3347945}.

After defining evaluation metrics, AR UI practitioners (e.g., software engineers and designers) often examine the dynamic user experience in interactive environments while satisfying the aforementioned metrics. There are attempts to assess the AR experience by building mini-size studio interactive spaces to emulate AR environments~\cite{m14-mini-10.1145/2702613.2732744}. However, such physical setups and iterative assessments are costly and time-consuming, especially when AR is implemented on large-scales (i.e., ubiquitously in our living spaces). Moreover, the increasing number of evaluation metrics calls for systematic evaluations of AR UIs~\cite{m12-10.1145/3282353.3282357}. It is therefore preferable to assess AR UI design metrics through systematic approaches and even automation, with prominent features of real-time monitoring of system performance, direct information collection via user-device interaction in AR, and more proactive responses to improve user perceptions to AR information~\cite{m12-10.1145/3282353.3282357}. However, to the best of our knowledge, the number of existing evaluation frameworks is very limited and their scopes are limited to specific contexts and scenarios, such as disaster management~\cite{m13-MRAT-10.1145/3313831.3376330}. More generic evaluation frameworks with a high selectivity of AR evaluation metrics pose research opportunities in the domain of AR interface designs~\cite{m3-10.1145/3385378.3385384, m12-10.1145/3282353.3282357}.
\section{MAR frameworks}\label{sec:framework}
MAR systems can either be deployed individually on client devices (e.g., smartphones and HMDs) without the need for separate hardware, or client devices and external servers together. 
This section investigates existing MAR frameworks, a large number of which are designed for the creation of MAR applications. A general hardware benchmark for these frameworks can be found by exploring the minimum requirements for MAR frameworks, such as ARCore and ARKit. These software development kits (SDKs) should be most optimised for the devices released by Google and Apple, respectively. According to the device-specification for both SDKs, the oldest devices still actively supported were released in 2015, i.e., the Nexus 5X \cite{google:arcore:devices} and the iPhone 6S \cite{apple:arkit}. However, older devices can still be supported by commercial and open-source solutions, such as ARToolKit which supports Android devices released in 2013 \cite{artoolkit:android_support}, and Vuforia which supports the 2012 released iPhone 5 \cite{vuforia:hardware_support}. Through software optimisation, MAR support for a variety of mobile hardware is possible. However, hardware backwards-compatibility can be reduced as more features are developed for the SDKs, and those features may require better processing capabilities or additional hardware, which may not always be available from older devices. Therefore, features may be omitted from version releases, or device support may be removed completely.  

\subsection{Evaluation Metrics} \label{ssec:framework_features}
MAR frameworks have various features as they aim to support different platforms and constraints. We compare these frameworks by evaluating the availability of the features, capabilities, and the devices which the frameworks can be deployed on as evaluation metrics. 

\begin{figure}
    \centering
    \includegraphics[width=0.88\textwidth]{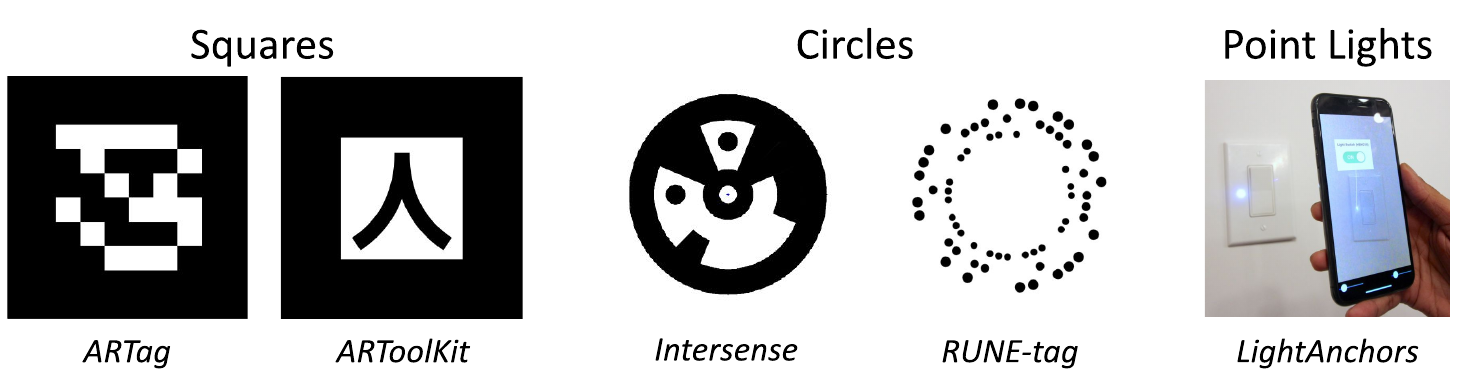}
    \caption{Examples of square, circular, and point light fiducial markers used in MAR.}
    \label{fig:tracking_examples}
\end{figure}

\begin{enumerate}
    \item \textbf{Platform support:}
    MAR is not necessarily restricted to only smartphones running Android (\faAndroid) or iOS (\faApple) operating systems (OSs). Wearables and HMDs are other MAR devices, which run alternative OSs, such as Linux (\faLinux) \cite{fragoso2011translatar, piekarski2001augmented} and Windows Holographic OS/Windows 10 (\faWindows) \cite{microsoft:hololens}. The frameworks are typically deployable on several different hardware types. Cross-platform support is also found through web browser-based MAR (\faHtml5) and using the cross-platform Unity Real-Time Development Platform (\unityIcon) \cite{kim2014using}.
    
    \item \textbf{Tracking:}
    To provide MAR experiences, virtual annotations are placed and tracked within physical environments. The two most commonly used techniques to enable and produce these experiences are 
    \begin{enumerate*}
        \item marker- or fiducial-based,
        and \item markerless tracking. 
    \end{enumerate*}
    \begin{itemize}
        \item Marker or fiducial tracking uses 2D patterns of predefined shapes and sizes, which are placed within environments, these are recognised from optical images and accompanying information associated with the patterns are returned to users \cite{fiala2005artag}. Among the different types of planar markers used in AR, Figure \ref{fig:tracking_examples} presents typical examples, including shaped-based markers such as squares (ARTag \cite{fiala2005artag}, ARToolKit \cite{artoolkit}), circles (Intersense \cite{naimark2002circular}, RUNE-tag \cite{bergamasco2016accurate}), and pervasive point lights (e.g., LEDs or light bulbs) to spatially-anchor data (LightAnchors \cite{ahuja2019lightanchors}). Planar patterns, such as QR codes \cite{kan2009applying} are good at encoding information. However, they do not function well in certain MAR situations such as when there are large fields of view and when perspectives are distorted \cite{fiala2005artag}.
        \item Markerless tracking utilises local environments of users to generate and fix virtual markers for 3D tracking \cite{ferrari2001markerless, neumann1999natural}. Also known as natural feature tracking (NFT), the natural features are points and regions of interest which are detected and selected, then motion and pose estimation are performed to track those features, and augmented content can be fixed to physical environments \cite{neumann1999natural}. The most popular techniques for this environment analysis is simultaneous localisation and mapping (SLAM) \cite{reitmayr2010simultaneous}. SLAM methods generally reconstruct an environment into a 3D spatial map (3D point cloud) to build a global reference map while simultaneously tracking the subject's position \cite{chen2017real, bailey2006simultaneous}. Markerless tracking has advantages such as not requiring prior knowledge of a user's locale, and not needing to place additional objects (e.g., fiducial markers) in an environment to create MAR experiences. 
    \end{itemize}

    Some frameworks also support a variety of other tracking techniques and behaviours which have varying degrees of support across the different frameworks \cite{arfoundation}, including
    \begin{enumerate*}
        \item Device tracking: the device's position and orientation are tracked with respect to physical space.
        \item Plane tracking: flat, horizontal and vertical surfaces are detected and tracked. 
        \item Hand tracking: human hands are recognised and tracked. 
        \item Body tracking: the entire human body is recognised in relation to physical space. 
        \item Facial tracking: the human face is detected and tracked.
    \end{enumerate*}

    \item \textbf{Features:} 
    Frameworks and SDKs support various features, which enhance the overall MAR experience for users, whether through quality of service, or quality of experience improvements \cite{arfoundation}:
    \begin{enumerate*}
        \item Point clouds: feature point maps from NFT, frameworks sometimes allow the maps to be accessible to developers and users.
        \item Anchors: these are points and locations of arbitrary position and orientation which the system tracks.
        \item Light estimation: the average colour temperature and brightness within a physical space is estimated. 
        \item Environment probes: probes which are used to generate a cube map which represents a particular area of a physical environment.
        \item Meshing: a physical space is virtually segmented and masked using triangular meshes. 
        \item Collaboration: positions and orientations of other devices are tracked and shared to create a collaborative MAR experience. 
        \item Occlusion: objects in the physical world have known or calculated distances to aid in the rendering of 3D content, which allows for realistic blending of physical and virtual worlds.
        \item Raycasting: the physical surroundings are queried for detected planes and feature points. 
        \item Pass-through video: the captured camera feed is rendered onto the device touchscreen as the background for MAR content.
        \item Session management: the platform-level configuration is changed automatically when AR features are enabled or disabled.
    \end{enumerate*}
    
    \item \textbf{Sensors:}
    The hardware providing MAR experiences are now equipped with a wide range of sensors. In addition, they contain high resolution displays and cameras; inertial measurement unit (IMU) sensors such as gyroscopes, accelerometers, and compasses; and Global Positioning System (GPS) sensors are now a standard inclusion in smartphones and HMDs. They benefit MAR frameworks with estimating user location and orientation to display location-relevant information \cite{carmigniani2011augmented}. Ranging technologies, such as LiDAR (light detection and ranging) sensors are now used to enhance current arrays of optical cameras on mobile smartphones for better spatial mapping, which is available from recent Apple iPhone and iPad devices \cite{apple:lidar}.
    
    \item \textbf{Architecture:}
    With every iteration release of mobile hardware, these devices are more capable with increasingly powerful CPUs and GPUs to perform onboard tasks (\textit{offline} processing). However, not every user has access to the latest and most powerful devices, therefore in these scenarios, a framework could employ external servers to offload computations (\textit{online} processing). Frameworks could then support either offline and online recognition of objects, or both. 
\end{enumerate}

Other considerations for selecting an MAR framework include, whether the framework is available for free, or requires the purchasing of a commercial software license; whether the framework is open source for developers to expand and to build upon; whether the framework is for general ``all-purpose'' usage, i.e., not just for one use-case like facial-tracking; and whether there are studio tools to simplify and aid in the development of MAR experiences.

\subsection{Frameworks}
We perform a comprehensive summary of existing MAR frameworks and SDKs, with considering the evaluation metrics specified in Section \ref{ssec:framework_features}. We tabulate our results in Table \ref{table:ar_frameworks}. To provide a basic filter on the frameworks and SDKs listed, we decide to exclude frameworks which have not been updated within a four year time span. This represents the age of the iPhone SE (1st generation, 2016) which is supported by the latest version of iOS 14 from Apple (i.e., the oldest supported iPhone) \cite{apple:ios14}, and similarly, Google supports the Pixel 2 (2017) in their latest Android 11 OS update \cite{google:android11}. Frameworks which last had updates four years ago may not function on the wide range of new hardware and software now available to users. Certainly, this may also be true for frameworks which have not had updates since one year ago, these are therefore marked in Table \ref{table:ar_frameworks} with an asterisk (*).

\begin{table}[t]
    \centering
    \tiny
    \renewcommand{\arraystretch}{1.3}
    \setlength\tabcolsep{1.94pt}  
    \caption{Comparisons of several available features in MAR frameworks and SDKs.}
    \label{table:ar_frameworks}
    \begin{tabularx}{\textwidth}{|l|cccccc|ccccccc|cccccccccc|cccc|ccccc|}
    \hline
    \multicolumn{1}{|l|}{\textit{\textbf{Framework/SDK}}} & \multicolumn{6}{l|}{\textit{\textbf{Platform support}}} & \multicolumn{7}{l|}{\textit{\textbf{Tracking}}} & \multicolumn{10}{l|}{\textit{\textbf{Features}}} & \multicolumn{4}{l|}{\textit{\textbf{Sensors}}} & \multicolumn{5}{l|}{\textit{\textbf{Others}}} \\
     
     & \faAndroid & \faApple & \faLinux & \faWindows & \faHtml5 & \unityIconTiny & \rotatebox[origin=c]{90}{Markers} & \rotatebox[origin=c]{90}{NFT} & \rotatebox[origin=c]{90}{Device} & \rotatebox[origin=c]{90}{Plane} & \rotatebox[origin=c]{90}{Hand} & \rotatebox[origin=c]{90}{2D \& 3D body} & \rotatebox[origin=c]{90}{Facial} & \rotatebox[origin=c]{90}{Point clouds} & \rotatebox[origin=c]{90}{Anchors} & \rotatebox[origin=c]{90}{Light estimation} & \rotatebox[origin=c]{90}{Environment probes} & \rotatebox[origin=c]{90}{Meshing} & \rotatebox[origin=c]{90}{Collaboration} & \rotatebox[origin=c]{90}{Occlusion} & \rotatebox[origin=c]{90}{Raycasting} & \rotatebox[origin=c]{90}{Pass-through video} & \rotatebox[origin=c]{90}{ Session management} & \rotatebox[origin=c]{90}{Camera} & \rotatebox[origin=c]{90}{LiDAR} & \rotatebox[origin=c]{90}{IMU} & \rotatebox[origin=c]{90}{GPS} & \rotatebox[origin=c]{90}{Architecture} & \rotatebox[origin=c]{90}{Price} & \rotatebox[origin=c]{90}{Open source} & \rotatebox[origin=c]{90}{General} & \rotatebox[origin=c]{90}{Studio tool} \\
    \hline
    
    A-Frame (1.2.0) \cite{aframe} & \ccy & \ccy & \ccy & \ccy & \ccy & \ccn & \ccn & \ccn & \ccn & \ccn & \ccn & \ccn & \ccn & \ccn & \ccn & \ccn & \ccn & \ccn & \ccn & \ccn & \ccn & \ccn & \ccn & \ccy & \ccn & \ccn & \ccn & Online & Free & \ccy & \ccy & \ccn \\ 
    \hline 
    
    ALVAR (0.7.2) \cite{alvar} & \ccy & \ccy & \ccy & \ccy & \ccn & \ccy & \ccy & \ccy & \ccy & \ccy & \ccn & \ccn & \ccn & \ccy & \ccy & \ccn & \ccn & \ccn & \ccn & \ccn & \ccn & \ccy & \ccn & \ccy & \ccn & \ccy & \ccy & Offline & Free & \ccn & \ccy & \ccn \\ 
    \hline
    
    Amazon Sumerian (N/A) \cite{amazonSumerian} & \ccy & \ccy & \ccn & \ccy & \ccy & \ccn & \ccy & \ccy & \ccy & \ccy & \ccn & \ccn & \ccn & \ccn & \ccy & \ccn & \ccy & \ccy & \ccn & \ccn & \ccy & \ccy & \ccn & \ccy & \ccn & \ccn & \ccn & Online & Free, Paid & \ccn & \ccy & \ccy \\
    \hline
    
    ApertusVR (0.9*) \cite{apertusvr} & \ccy & \ccn & \ccn & \ccy & \ccn & \ccn & \ccn & \ccn & \ccn & \ccn & \ccn & \ccn & \ccn & \ccn & \ccn & \ccn & \ccy & \ccn & \ccn & \ccn & \ccy & \ccy & \ccy & \ccy & \ccn & \ccn & \ccn & Offline & Free & \ccy & \ccy & \ccn \\
    \hline
    
    ARCore (1.23.0) \cite{google:arcore} & \ccy & \ccy & \ccn & \ccn & \ccn & \ccy & \ccy & \ccy & \ccy & \ccy & \ccn & \ccn & \ccy & \ccy & \ccy & \ccy & \ccy & \ccn & \ccy & \ccy & \ccy & \ccy & \ccy & \ccy & \ccn & \ccy & \ccy & Offline & Free & \ccy & \ccy & \ccn \\
    \hline
    
    ARKit (4) \cite{apple:ar} & \ccn & \ccy & \ccn & \ccn & \ccn & \ccy & \ccy & \ccy & \ccy & \ccy & \ccy & \ccy & \ccy & \ccy & \ccy & \ccy & \ccy & \ccy & \ccy & \ccy & \ccy & \ccy & \ccy & \ccy & \ccy & \ccy & \ccy & Offline & Free & \ccn & \ccy & \ccy \\
    \hline
    
    ARMedia SDK (2.1.0*) \cite{armediasdk} & \ccy & \ccy & \ccn & \ccy & \ccn & \ccy & \ccy & \ccy & \ccy & \ccy & \ccn & \ccn & \ccn & \ccn & \ccn & \ccn & \ccn & \ccn & \ccn & \ccn & \ccn & \ccy & \ccn & \ccy & \ccn & \ccy & \ccy & Offline & Paid & \ccn & \ccy & \ccn \\
    \hline
    
    ARToolKit (5.4*) \cite{artoolkit, kato1999marker} & \ccy & \ccy & \ccy & \ccy & \ccn & \ccn & \ccy & \ccn & \ccn & \ccy & \ccn & \ccn & \ccn & \ccn & \ccn & \ccn & \ccn & \ccn & \ccn & \ccn & \ccn & \ccy & \ccn & \ccy & \ccn & \ccn & \ccn & Offline & Free & \ccy & \ccy & \ccn \\
    \hline
    
    artoolkitX (1.0.6.1) \cite{artoolkitx:ar} & \ccy & \ccy & \ccy & \ccy & \ccn & \ccn & \ccy & \ccn & \ccn & \ccn & \ccn & \ccn & \ccn & \ccn & \ccn & \ccn & \ccn & \ccn & \ccn & \ccn & \ccn & \ccy & \ccn & \ccy & \ccn & \ccn & \ccn & Offline & Free & \ccy & \ccy & \ccn \\
    \hline
    
    ArUco (3.1.12) \cite{aruco, ROMERORAMIREZ201838} & \ccy & \ccn & \ccy & \ccy & \ccn & \ccn & \ccy & \ccn & \ccy & \ccy & \ccn & \ccn & \ccn & \ccn & \ccn & \ccn & \ccn & \ccn & \ccn & \ccn & \ccn & \ccy & \ccn & \ccy & \ccn & \ccn & \ccn & Offline & Free & \ccy & \ccy & \ccn \\
    \hline
    
    AR.js (3.3.1) \cite{ar.js:ar} & \ccy & \ccy & \ccy & \ccy & \ccy & \ccn & \ccy & \ccn & \ccn & \ccn & \ccn & \ccn & \ccn & \ccn & \ccn & \ccn & \ccn & \ccn & \ccn & \ccn & \ccn & \ccy & \ccn & \ccy & \ccn & \ccn & \ccy & Online & Free & \ccy & \ccy & \ccn \\
    \hline
    
    Augment (4.0.6*) \cite{augment} & \ccy & \ccy & \ccn & \ccn & \ccy & \ccn & \ccy & \ccy & \ccn & \ccn & \ccn & \ccn & \ccn & \ccn & \ccn & \ccn & \ccn & \ccn & \ccn & \ccn & \ccn & \ccy & \ccy & \ccy & \ccn & \ccn & \ccn & Online & Paid & \ccn & \ccn & \ccy  \\
    \hline
    
    Augmented Pixels (N/A) \cite{augmentedPixels} & \ccy & \ccy & \ccn & \ccn & \ccn & \ccn & \ccn & \ccy & \ccn & \ccn & \ccn & \ccn & \ccn & \ccy & \ccy & \ccn & \ccn & \ccn & \ccn & \ccn & \ccn & \ccy & \ccn & \ccy & \ccn & \ccy & \ccy & Offline & Paid & \ccn & \ccy & \ccn \\
    \hline
    
    AugmentedPro (2.4.3) \cite{augmentedPro} & \ccy & \ccn & \ccn & \ccy & \ccn & \ccn & \ccy & \ccn & \ccn & \ccn & \ccn & \ccn & \ccn & \ccn & \ccn & \ccn & \ccn & \ccn & \ccn & \ccn & \ccn & \ccy & \ccn & \ccy & \ccn & \ccn & \ccn & Offline & Paid & \ccn & \ccy & \ccy \\
    \hline
    
    Banuba (0.35.0) \cite{banuba} & \ccy & \ccy & \ccn & \ccn & \ccy & \ccy & \ccn & \ccy & \ccn & \ccn & \ccn & \ccn & \ccy & \ccn & \ccn & \ccy & \ccn & \ccy & \ccn & \ccy & \ccn & \ccy & \ccn & \ccy & \ccn & \ccy & \ccn & Offline & Paid & \ccn & \ccn & \ccn \\
    \hline
    
    Blippar (N/A) \cite{blippAR} & \ccy & \ccy & \ccn & \ccn & \ccy & \ccn & \ccy & \ccy & \ccy & \ccy & \ccn & \ccn & \ccy & \ccy & \ccn & \ccn & \ccn & \ccn & \ccn & \ccn & \ccn & \ccy & \ccy & \ccy & \ccn & \ccy & \ccy & Online & Paid & \ccn & \ccy & \ccy \\
    \hline
    
    CraftAR (5.2.1*) \cite{craftar:ar} & \ccy & \ccy & \ccn & \ccn & \ccn & \ccy & \ccy & \ccn & \ccy & \ccn & \ccn & \ccn & \ccn & \ccn & \ccy & \ccn & \ccn & \ccn & \ccn & \ccy & \ccn & \ccy & \ccy & \ccy & \ccn & \ccn & \ccn & Both & Free, Paid & \ccn & \ccy & \ccy \\
    \hline
    
    DeepAR (2.3.1) \cite{deepar} & \ccy & \ccy & \ccn & \ccn & \ccy & \ccy & \ccn & \ccy & \ccn & \ccn & \ccn & \ccn & \ccy & \ccn & \ccn & \ccn & \ccn & \ccy & \ccn & \ccy & \ccn & \ccy & \ccn & \ccy & \ccn & \ccn & \ccn & Offline & Free, Paid & \ccn & \ccn & \ccy \\
    \hline
    
    EasyAR (4.2.0) \cite{easyar} & \ccy & \ccy & \ccn & \ccy & \ccn & \ccy & \ccy & \ccy & \ccy & \ccy & \ccn & \ccn & \ccn & \ccy & \ccn & \ccn & \ccn & \ccy & \ccn & \ccy & \ccy & \ccy & \ccy & \ccy & \ccn & \ccy & \ccn & Offline & Free, Paid & \ccn & \ccy & \ccn \\
    \hline
    
    HERE SDK (3.17) \cite{heresdk} & \ccy & \ccy & \ccn & \ccn & \ccn & \ccn & \ccn & \ccy & \ccy & \ccn & \ccn & \ccn & \ccn & \ccn & \ccy & \ccn & \ccn & \ccn & \ccn & \ccy & \ccn & \ccy & \ccy & \ccy & \ccn & \ccy & \ccy & Offline & Paid & \ccn & \ccn & \ccn \\
    \hline
    
    Kudan AR SDK (1.6.0) \cite{kudan_ar} & \ccy & \ccy & \ccn & \ccn & \ccn & \ccy & \ccy & \ccy & \ccy & \ccy & \ccn & \ccn & \ccn & \ccy & \ccn & \ccn & \ccn & \ccn & \ccn & \ccn & \ccn & \ccy & \ccn & \ccy & \ccn & \ccy & \ccy & Offline & Free, Paid & \ccn & \ccy & \ccn \\
    \hline
    
    Lumin SDK (0.25.0) \cite{lumin} & \ccn & \ccn & \ccn & \ccn & \ccn & \ccy & \ccy & \ccy & \ccy & \ccy & \ccy & \ccn & \ccn & \ccy & \ccy & \ccn & \ccn & \ccy & \ccy & \ccy & \ccy & \ccn & \ccy & \ccy & \ccn & \ccy & \ccn & Offline & Free & \ccn & \ccy & \ccn \\
    \hline
    
    MAXST AR SDK (5.0.3) \cite{maxst} & \ccy & \ccy & \ccn & \ccy & \ccn & \ccy & \ccy & \ccy & \ccy & \ccy & \ccn & \ccn & \ccn & \ccy & \ccn & \ccn &  \ccn & \ccy & \ccn & \ccy & \ccn & \ccy & \ccn & \ccy & \ccn & \ccy & \ccn & Both & Free, Paid & \ccn & \ccy & \ccn \\
    \hline
    
    Minsar (2.0) \cite{minsar} & \ccy & \ccy & \ccn & \ccn & \ccy & \ccn & \ccy & \ccy & \ccy & \ccy & \ccn & \ccn & \ccn & \ccn & \ccy & \ccy & \ccn & \ccn & \ccn & \ccy & \ccn & \ccy & \ccy & \ccy & \ccn & \ccy & \ccy & Both & Free, Paid & \ccn & \ccy & \ccy \\
    \hline
    
    MRTK (2.6.0) \cite{mrtk} & \ccy & \ccy & \ccn & \ccy & \ccn & \ccy & \ccy & \ccy & \ccn & \ccy & \ccy & \ccn & \ccn & \ccn & \ccy & \ccn & \ccn & \ccy & \ccy & \ccy & \ccn & \ccn & \ccn & \ccy & \ccn & \ccy & \ccn & Offline & Free & \ccy & \ccy & \ccn \\
    \hline
    
    NyARToolkit (5.0.8*) \cite{nyartoolkit} & \ccy & \ccn & \ccn & \ccy & \ccn & \ccy & \ccy & \ccy & \ccn & \ccn & \ccn & \ccn & \ccn & \ccn & \ccn & \ccn & \ccn & \ccn & \ccn & \ccn & \ccn & \ccy & \ccn & \ccn & \ccn & \ccn & \ccn & Offline & Free & \ccy & \ccy & \ccn \\
    \hline
    
    Onirix (N/A) \cite{onirix} & \ccy & \ccy & \ccn & \ccn & \ccn & \ccy & \ccy & \ccy & \ccn & \ccy & \ccn & \ccn & \ccn & \ccy & \ccy & \ccy & \ccn & \ccn & \ccn & \ccn & \ccy & \ccy & \ccy & \ccy & \ccn & \ccy & \ccy & Online & Paid & \ccn & \ccy & \ccy \\
    \hline
    
    Pikkart AR SDK (3.5.8*) \cite{pikkart} & \ccy & \ccy & \ccn & \ccn & \ccn & \ccy & \ccy & \ccn & \ccy & \ccy & \ccy & \ccn & \ccn & \ccn & \ccn  & \ccn & \ccn & \ccn & \ccn & \ccn & \ccn & \ccy & \ccy & \ccy & \ccy & \ccy & \ccy & Both & Free, Paid & \ccn & \ccy & \ccn \\
    \hline
    
    PlugXR (1.0.0) \cite{plugxr} & \ccy & \ccy & \ccy & \ccy & \ccy & \ccy & \ccy & \ccy & \ccn & \ccn & \ccn & \ccn & \ccn & \ccn & \ccn & \ccn & \ccn & \ccn & \ccn & \ccn & \ccn & \ccy & \ccy & \ccy & \ccn & \ccn & \ccn & Online & Paid & \ccn & \ccy & \ccy \\
    \hline
    
    Universal AR SDK (N/A) \cite{universalarsdk} & \ccy & \ccy & \ccy & \ccy & \ccy & \ccy & \ccy & \ccy & \ccn & \ccy & \ccn & \ccn & \ccy & \ccn & \ccy & \ccn & \ccn & \ccn & \ccn & \ccn & \ccn & \ccy & \ccn & \ccy & \ccn & \ccn & \ccn & Online & Free & \ccn & \ccy & \ccn \\
    \hline
    
    Vectary (N/A) \cite{vectary} & \ccy & \ccy & \ccy & \ccy & \ccy & \ccn & \ccn & \ccy & \ccn & \ccy & \ccn & \ccn & \ccn & \ccn & \ccn & \ccn & \ccn & \ccn & \ccn & \ccn & \ccn & \ccy & \ccn & \ccy & \ccn & \ccn & \ccn & Online & Paid & \ccn & \ccn & \ccy \\
    \hline
    
    Vidinoti SDK (N/A) \cite{vidinoti} & \ccy & \ccy & \ccn & \ccn & \ccn & \ccn & \ccy & \ccn & \ccy & \ccy & \ccn & \ccn & \ccn & \ccn & \ccn & \ccn & \ccn & \ccn & \ccn & \ccn & \ccn & \ccy & \ccn & \ccy & \ccn & \ccn & \ccy & Both & Paid & \ccn & \ccy & \ccy \\
    \hline
    
    ViewAR SDK (N/A) \cite{viewarsdk} & \ccy & \ccy & \ccn & \ccn & \ccn & \ccn & \ccy & \ccy & \ccn & \ccn & \ccn & \ccn & \ccn & \ccy & \ccn & \ccn & \ccn & \ccn & \ccy & \ccn & \ccn & \ccy & \ccy & \ccy & \ccn & \ccy & \ccy & Offline & Paid & \ccn & \ccy & \ccn \\
    \hline
    
    Vuforia (9.7) \cite{ptc:vuforia} & \ccy & \ccy & \ccn & \ccy & \ccn & \ccy & \ccy & \ccy & \ccy & \ccy & \ccn & \ccn & \ccn & \ccn & \ccy & \ccy & \ccn & \ccn & \ccn & \ccn & \ccy & \ccy & \ccy & \ccy & \ccn & \ccn & \ccy & Both & Free, Paid & \ccn & \ccy & \ccn \\
    \hline
    
    Wikitude (9.6) \cite{wikitude:ar} & \ccy & \ccy & \ccn & \ccy & \ccn & \ccy & \ccy & \ccy & \ccy & \ccy & \ccn & \ccn & \ccn & \ccy & \ccy & \ccn & \ccn & \ccn & \ccn & \ccy & \ccy & \ccy & \ccy & \ccy & \ccn & \ccy & \ccy & Both & Paid & \ccn & \ccy & \ccy \\ 
    \hline
    
    WebXR (N/A) \cite{baruahar} & \ccy & \ccy & \ccn & \ccy & \ccy & \ccn &
    \ccy & \ccy & \ccy & \ccy & \cco & \cco& \cco & 
    \cco & \ccy & \ccy & \ccy & \ccy & \ccn & \ccy & \ccy & \ccy & \ccy &
    \ccy & \cco & \cco & \cco & Offline & Free & \ccy & \ccy & \ccn \\ 
    \hline
    
    XZIMG (2.0.2*) \cite{xzimg} & \ccy & \ccy & \ccy & \ccy & \ccy & \ccy & \ccy & \ccy & \ccn & \ccy & \ccy & \ccn & \ccy & \ccn & \ccn & \ccn & \ccn & \ccn & \ccn & \ccn & \ccn & \ccy & \ccn & \ccy & \ccn & \ccn & \ccn & Offline & Free, Paid & \ccn & \ccy & \ccn \\
    \hline
    \end{tabularx}
    \begin{tablenotes}
        \tiny
        \item Frameworks and SDKs marked with an asterisk (*) have not been updated in $1+$ years. 
        \item Features or functions optionally supported in different platform are marked with \ding{109}.  
    \end{tablenotes}
\end{table}
\begin{itemize}
    \item \textbf{A-Frame} \cite{aframe} is a framework primarily dedicated for web-based virtual reality (VR). Built using JavaScript, A-Frame has additional capabilities for displaying AR content. However, greater MAR functionality is provided when using the community component, AR.js.   

    \item \textbf{ALVAR} \cite{alvar} is a suite of SDKs developed by the VTT Technical Research Centre in Finland. Functionality is split into three distinct C++ libraries: \begin{enumerate*} \item \textit{ALVAR Core} is a low-level AR toolkit; \item \textit{ALVAR Platform} abstracts the camera, IMU, and GPS sensors and a data processing pipeline; and \item \textit{ALVAR Tracker} provides 3D point cloud, planar image, multi-marker, and panorama tracking. \end{enumerate*}
    
    \item \textbf{Amazon Sumerian} \cite{amazonSumerian} is an Amazon-managed service for building and deploying AR, VR, and 3D applications. Sumerian applications are accessible through web browsers and is based on JavaScript API and WebGL 2. Additionally, hybrid experiences can be achieved with iOS ARKit and Android ARCore. 
    
    \item \textbf{ApertusVR} \cite{apertusvr} is primarily an open source VR software library composed of a main core (ApertusCore) and a set of plugins (ApertusVR) which extends the core to provide both AR and VR functionality. Their architecture is based on a ``distributed plugin mechanism'' which allows for IoT devices and other smart hardware to interface with the system.

    \item \textbf{ARCore} \cite{google:arcore} is an SDK developed by Google for mobile platforms, with the main development environment being Android smartphones. However, there are APIs which enable cross-platform AR experiences. One notable service is ARCore Cloud Anchor which allows for anchors and feature points to be sent to a cloud host, and subsequently shared with multiple user devices to create collaborative and multiplayer apps. 
    
    \item \textbf{ARKit} \cite{apple:ar} is Apple's solution to MAR for their products, i.e., iPhone, iPad, and iPod devices. Visual-inertial odometry is used to analyse notable features from a scene taken by a device's camera to provide physical world tracking, and there are additional classes to track people's movement using body anchors, and using alpha and depth information from an image frame to occlude virtual content. The framework can additionally utilise LiDAR (light detection and ranging) data from an iPhone and iPad sensor to enhance spatial environment mapping.   
    
    \item \textbf{ARMedia SDK} \cite{armediasdk} is developed by Inglobe Technologies which provides object, planar image, geolocation-based and motion tracking. Additionally, a simple object rendering engine is included to visualise 3D contents.
    
    \item \textbf{ARToolKit} \cite{artoolkit, kato1999marker} was a software library developed by the Human Interface Technology Laboratory at the University of Washington. The framework uses camera position and orientation relative to physical fiducial markers to determine scene geometry. The last update to the original ARToolKit SDK was made in May 2007 \cite{artoolkit:sourceforge}. Android, iOS and JavaScript support was later introduced in ARToolKit in 2015 when version 5 was released as an open-source framework by DAQRI \cite{artoolkit:github}. This project was developed until November 2017, when the developers' focus shifted to a new open-source project, artoolkitX.
    
    \item \textbf{artoolkitX} \cite{artoolkitx:ar} is a framework supported by Realmax Inc. and has four distinct libraries: \begin{enumerate*} \item camera and video source processing, \item feature point tracking with OpenCV, \item fiducial marker tracking, and \item virtual content rendering with OpenGL/GLES \end{enumerate*}. All of these components are managed by a central controller class.
    
    \item \textbf{ArUco} \cite{aruco, ROMERORAMIREZ201838} is a minimal open source library for camera pose estimation using square markers. Discrimination correlation filters are used for the tracking of markers and their corners, and achieves continuous detection under extreme blurring and scaling conditions. 
    
    \item \textbf{AR.js} \cite{ar.js:ar} is an MAR library built for mobile smartphone web browsers. Adapting the 2015 JavaScript version of ARToolKit, this framework is a lightweight AR solution which uses fiducial markers and on-device GPS sensors to localise and track 3D content. The only requirement client devices must have, is the web browser should contain support for WebGL and WebRTC JavaScript APIs. Users do not need to install a device application, rather, just load a web app. AR.js can be expanded with additional JavaScript frameworks such as A-Frame \cite{aframe} and three.js \cite{threejs} to provide alternative methods for rendering virtual content.
    
    \item \textbf{Augment} \cite{augment} is a paid platform for 3D and AR product visualisation for mobile devices and web browsers. The platform has several products for viewing and managing 3D models, and an SDK to embed the model player in custom applications.
    
    \item \textbf{Augmented Pixels} \cite{augmentedPixels} is an SDK which provides a custom SLAM platform for developers to fuse data from multiple sensor inputs to calculate positioning. The primary deployment scenarios are drones and robotics navigation, and providing real-time SLAM for mobile smartphones and AR/VR glasses.
    
    \item \textbf{AugmentedPro} \cite{augmentedPro} is a two component AR solution for industry. An \textit{AugmentedPro Creator} program is used to modify and create virtual content, this can then be viewed and displayed as an AR experience through a companion \textit{AugmentedPro Player}.
    
    \item \textbf{Banuba} \cite{banuba} provides several augmented face AR solutions. They have two main SDKs, the first of which is tracking and augmenting facial features in a Face AR SDK, the second is an AI Video Editor SDK which allows users to alter their video streams through the inclusion of face filters, modifying video backgrounds, and applying AR video overlays. 
    
    \item \textbf{Blippar} \cite{blippAR} provides several products for commercial and educational AR applications. Developers can develop their own experiences in custom applications using the Blippar AR SDK. Alternatively, their Blippar app and their creation tools allows for a simplified development process, as their application functions as an AR browser to which users can obtain multiple AR experiences for different products and services. The Blippar SDK is also used for other frameworks such as Layar \cite{layar}. 
    
    \item \textbf{CraftAR} \cite{craftar:ar} is a marker-based AR SDK providing content rendering on-device or through their cloud service. There is an image recognition solution which is also marketed under the same name \cite{craftar:image_recognition}, and provides on-device or cloud-based recognition. Both SDKs are combined and utilised as part of their Pro SDK product.
    
    \item \textbf{DeepAR} \cite{deepar} is an AR platform mainly focusing on providing Snapchat- and Instagram-style face augmentations. Utilising neural networks and deep learning models, DeepAR is able to perform face detection, face tracking, and emotion recognition. Face shape heatmaps and facial feature points are generated and used to track facial features.
    
    \item \textbf{EasyAR} \cite{easyar} is an SDK supporting surface, 3D object, and planar image tracking; as well as motion tracking by utilising visual inertial odometry, and on-device 3D mesh generation to create sparse and dense environment spatial maps. This allows for features such as persistent AR content, collaborative multi-user AR, collision detection, and object occlusion effects. 
    
    \item \textbf{HERE SDK} \cite{heresdk} is primarily a collection of SDKs and APIs which are used to build location-aware apps such as navigation applications. The main AR component of both SDKs for Android and iOS is the LiveSight feature which enables geospatial location markers in the real world. 
    
    \item \textbf{Kudan AR SDK} \cite{kudan_ar} is developed by Kudan, and is supported by KudanSLAM, Kudan's own SLAM solution. Their implementation uses hardware sensors such as stereo and visual-inertial depth cameras to create MAR experiences.
    
    \item \textbf{Lumin SDK} \cite{lumin} is the SDK by Magic Leap for producing native apps which are deployed on their AR smartglasses product, the Magic Leap 1. The SDK is available for several development platforms, and allows developers to create their self-described ``spatial computing'' experiences. 
    
    \item \textbf{MAXST AR SDK} \cite{maxst} is a cross-platform AR engine for developing MAR applications. Visual SLAM is used to obtain a map of the virtual environment, and subsequent coordinate systems are created with the \textit{MAXST AR Fusion Tracker} which uses technology from ARKit and ARCore.  
    
    \item \textbf{Minsar}, or Minsar Studio \cite{minsar} is a ``no-code'' tool for creating both AR and VR content for mobile devices, and HMDs (i.e., HoloLens, Magic Leap, and Oculus Quest). MAR experiences are made using the Minsar Studio application found on both iOS and Android, and are exported using either the companion \textit{XR View} application, or through a web browser using WebXR.    
    
    \item \textbf{MRTK}, or Mixed Reality Toolkit \cite{mrtk} is a Microsoft-driven framework for cross-platform mixed reality (MR) application development. MRTK primarily supports and drives applications on Microsoft HoloLens devices, but supports OpenVR headsets, and mobile devices running iOS and Android. On HoloLens 2 devices, MRTK provides additional features such as hand and eye tracking.
    
    \item \textbf{NyARToolkit} \cite{nyartoolkit} is a collection of AR libraries based on ARToolKit. Several platforms are supported including Android and Unity, where both marker-based and NFT tracking is available to serve augmented content. 
    
    \item \textbf{Onirix} \cite{onirix} is composed of three SDKs: \begin{enumerate*} \item \textit{Onirix Places} utilises user location and viewing direction along with Google Maps to provide geo-spatial AR content, \item \textit{Onirix Spaces} employs SLAM algorithms to provide markerless AR in indoor environments, and \item \textit{Onirix Targets} uses 2D marker objects and images for marker-based AR \end{enumerate*}. Onirix uses ARCore and ARKit to deliver the SLAM and light estimation features within the Onirix Spaces SDK.
    
    \item \textbf{Pikkart AR SDK} \cite{pikkart} utilises marker images and geolocation data to track virtually drawn content. A notable feature is Pikkart AR Logo, which uses printable steganography to produce micro-variations on markers so different virtual content can be returned using the ``same'' marker. 
    
    \item \textbf{PlugXR} \cite{plugxr} is a cloud-based MAR solution containing a \textit{PlugXR Creator} tool to create MAR experiences which are deployable on several platforms, and a \textit{PlugXR App} to view those experiences. The primary focus is providing tools so users with no coding knowledge can create AR applications. 
    
    \item \textbf{Universal AR SDK} \cite{universalarsdk} by ZapWorks is composed of several SDKs for different platforms to enable MAR experiences, including web browser-based MAR through compatibility with JavaScript, A-Frame, and ThreeJS. The main features across the frameworks include image, world, and face tracking.   
    
    \item \textbf{Vectary} \cite{vectary} is a web-based studio tool to create 3D and AR experiences. Those experiences are then accessible through mobile or desktop web browsers using their Vectary Web AR system.
    
    \item \textbf{Vidinoti SDK} \cite{vidinoti} is an SDK which delivers MAR experiences using a combination of on-device image recognition and tracking, as well as their cloud recognition service, however, their SDK is able to function completely offline.
    
    \item \textbf{ViewAR SDK} \cite{viewarsdk} allows developers to create both AR and VR applications using a JavaScript-based API. Both ARCore and ARKit frameworks are integrated within the SDK and are automatically selected depending on the device type. 
    
    \item \textbf{Vuforia} \cite{ptc:vuforia} is an SDK which can use several different types of images and objects as targets: pre-existing 3D models, planar images, cylindrical objects (e.g., columns, coffee cups, beverage bottles/cans), and targets can be created through scanning objects to produce 3D models. Additionally, video data from external non-client cameras can be used to enhance the recognition and tracking capabilities of the platform. 
    
    \item \textbf{Wikitude} \cite{wikitude:ar} is an MAR orientated SDK. Offering their own SLAM technology which can connect with ARCore and ARKit backends, and supports object fixing onto a scene even after the target has left the field of view. Traditional marker-based recognition is also available in the SDK, along with location-awareness and model occlusion. Wikitude is also deployable and optimised for smart glasses such as Epson Moverio and HoloLens.
    
    \item \textbf{WebXR} \cite{baruahar} is a specification developed by the World Wide Web Consortium, W3C. WebXR API is an implementation of the WebXR feature set. Since the WebXR API serves as an interface between XR Web content and the devices on which they run, the functions and features are not the same across different platforms or web browsers. 
    
    \item \textbf{XZIMG} \cite{xzimg} produces two main AR products, \textit{Augmented Vision} and \textit{Augmented Face}. Both are deployable on standard MAR hardware, as well as through web browsers. Augmented Vision is a typical fiducial marker recognition system, while Augmented Face has face tracking and augmentation technology. 
\end{itemize}

From the 37 listed frameworks, Android and iOS are the two most consistently supported platforms. Support for Unity is also often available as it simplifies the MAR application development process, e.g., easily exporting produced applications to multiple platforms. In terms of support for different tracking technologies: 
\begin{enumerate*}
    \item 30 of the 37 frameworks support 2D image marker tracking,
    \item 27 support natural feature tracking,
    \item only 6 support hand tracking,
    \item facial tracking is found in 8 of the frameworks,
    and \item only 2 frameworks supports 2D and 3D body tracking.
\end{enumerate*}
This indicates that there is still a primary focus on environment-based MAR, and human body-based tracking is still of limited interest to framework developers.

By comparison, there is varied supported for the different features in frameworks. Both ARCore and ARKit support the majority of the described features, they therefore form the baselines upon which other frameworks can be compared to and built upon. Among the 10 listed features, the following are organised as the most-to-least supported:
\begin{enumerate*}
    \item pass-through video,
    \item session management,
    \item anchors,
    \item point clouds and occlusion,
    \item raycasting,
    \item meshing,
    \item light estimation,
    and \item collaboration and environmental probing.
\end{enumerate*}
While the overall average number of supported features is approximately 3 per framework, there are 10 frameworks which only support 1, i.e., pass-through video. In relation to the type of architecture in use, offline and on-device-based processing is the most popular with 28 frameworks (21 offline-based, and 7 supporting both offline and online), and 16 frameworks use purely an online-based processing architecture. These online frameworks are often obtainable as paid services which are offered alongside a free trial variant, while the offline frameworks are generally free, albeit not always open source.
\section{Machine Learning Methods for MAR} \label{sec:mlmethods}
Different MAR frameworks and implementations contain approximately the same basic components, irrespective of the hardware used, or whether the system is self-contained on one device or distributed between client and an external server. Figure \ref{fig:mar_pipeline} in Section \ref{sec:introduction} presents the typical pipeline of MAR systems where different tasks function together to create the MAR experience. Several of these tasks often require the use of ML algorithms to be successfully completed, namely, machine vision-based tasks. ML is a set of methods and techniques used for modelling or building programs that employ knowledge gathered from solving previous problems to answer unseen examples of similar problems, i.e., learning from past experiences \cite{miraftabzadeh2019survey, michie1994machine}. ML technologies are crucial for MAR adoption, because they make the processing of visual, audio, and other sensor data to be more intelligent while simultaneously protecting privacy and security. A notable subfield of ML is deep learning, where multiple layers are used to represent the abstractions of data to build computational models \cite{pouyanfar2018survey}. Deep learning methods are useful in different domains \cite{abiodun2018state}. 

\begin{figure} [t]
    \centering
    \includegraphics[width=0.80\textwidth]{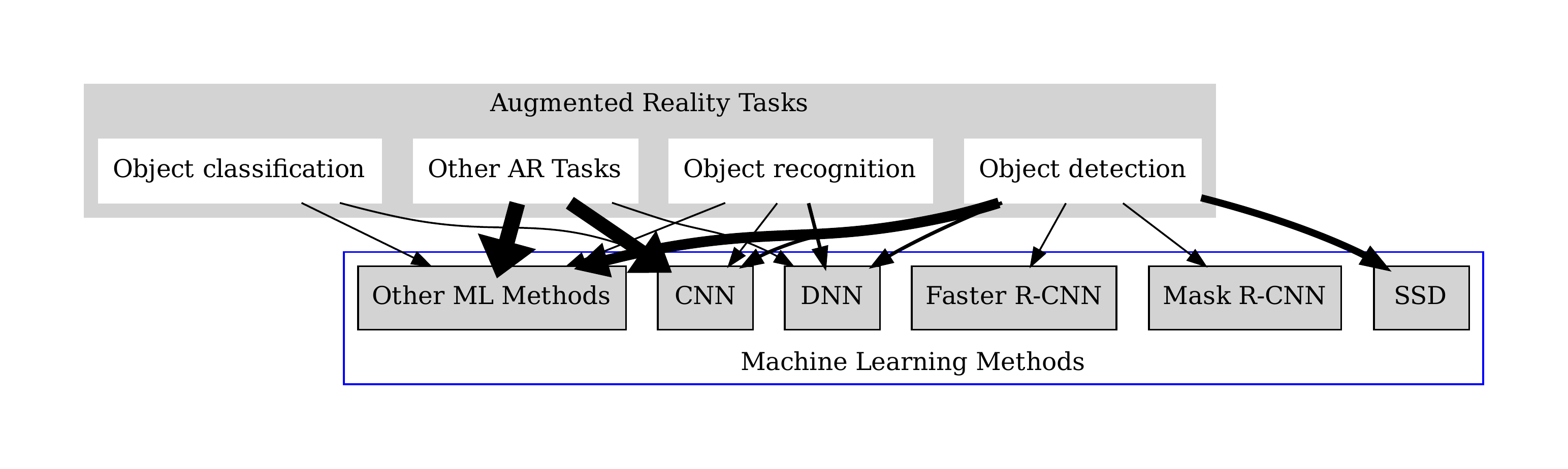}
    \caption{Mapping of AR tasks found in the works of Table \ref{tab:ml_methods} to the ML methods used to fulfil the tasks. The width thickness of the arrow lines denotes the amount of works associated with the tasks-to-ML method.}
    \label{fig:arml_map}
\end{figure}

In this section, we explore several MAR tasks which require the usage of ML methods. More specifically, we provide the requirements of the tasks, the ML methods used to fulfil the tasks, and the advantages and disadvantages of those methods. Table \ref{tab:ml_methods} provides a collection of most significant work of the aforementioned AR tasks as well as the ML methods used to fulfil them based on where ML algorithms are deployed on, Server, Edge, or Client. Subsequently, Figure \ref{fig:arml_map} provides a summary mapping of these tabulated works, linking the AR tasks to the ML methods, this figure demonstrates the popularity of certain methods, such as CNN and SSD (more details in Section \ref{ssec:ml_objdet}), and the roles they play in accomplishing AR tasks. 

\begin{table}
 \tiny
 \centering
 \caption{List of example AR works which use ML methods for completing AR tasks or as part of the system.}
  \label{tab:ml_methods}
    \tiny
    \begin{tabular}{|p{0.8cm}|p{5.5cm}|p{3.5cm}|p{3cm}|}
    \hline
    Device & Overview & Task & ML method  \\
    \hline
    \multirow{28}{*}{Server} & Deep learning for AR tracking \cite{akgul2016applying} & Object detection & CNN \\
    
    & Real-time moving object detection for AR \cite{cuevas2012moving} & Object detection & Background-foreground non-parametric-based \\
    
    & Deep-learning-based smart task assistance for wearable AR (HoloLens) \cite{park2020deep} & Object detection and instance segmentation & Mask R-CNN \\
    
    & Interacting with IoT devices in AR environment\cite{sun2019magichand} & Real-time hand gesture recognition & (2D) CNN \\
    
    & Edge-assisted distributed DNN for mobile WebAR \cite{ren2020edge} & Object recognition & DNN \\
    
    & Object detection and tracking for face and eyes AR \cite{hbali2013object} & Object detection, object recognition & History of Oriented Gradientsm, Haar-like features \\
    
    & AR surgical scene understanding improved with ML \cite{pauly2015machine} & Object identification & Random forest \\
    
    & Dynamic image recognition methods for AR \cite{cheng2020augmented} & Feature extraction, object recognition & CNN, XGBoost \\

    & AR platform for interactive aerodynamic design and analysis \cite{badias2019augmented} & Object recognition & Manifold learning \\ 
    
    & AR retail product identification \cite{upadhyay2020augmented} & Object detection & SSD \\
    
    & Improving retail shopping experience with AR \cite{cruz2019augmented} & Object recognition & ResNet50 (CNN) \\
    
    & AR instructional system for mechanical assembly \cite{lai2020smart} & Object detection & Faster R-CNN \\
    
    & Deep learning for AR \cite{lalonde2018deep} & Object tracking, light estimation & CNN \\
    
    & AR for radiology \cite{trestioreanu2020using} & Image segmentation & CNN \\
    
    & AR design personalisation for facial accessory products \cite{huang2012human} & Facial tracking & AdaBoost \\
    
    & Low cost AR for automotive industry \cite{ramasubramaniam2018lcar} & Feature extraction, object classification & Linear SVM, CNN \\
    
    & AR training framework for neonatal endotracheal intubation \cite{zhao2020intelligent} & Assessing task performance & CNN \\
    
    & AR video calling with WebRTC API \cite{jikadra2019video} & Semantic segmentation & CNN \\
    
    & AR gustatory manipulation \cite{nakano2019deeptaste} & Food-to-food translation & GAN \\
    
    \hline
    \multirow{9}{*}{Edge} & Edge-based inference for ML at Facebook \cite{wu2019machine} & Image classification & DNN \\
    
    & Supporting vehicle-to-edge for vehicle AR \cite{zhou2019enhanced} & Object detection & Deep CNN (YOLO) \\
    
    & AR platform for operators in production environments \cite{um2018modular} & Object detection & SSD \\
    
    & Spatial AR with single IR camera \cite{hashimoto2016dynamic} & 3D pose estimation, image classification & Hough Forests, Random Ferns \\
    
    & Federated learning for low-latency object detection and classification \cite{chen2020federated} & Object classification modelling & Federated learning \\
    
    & Optimising learning accuracy in MAR systems \cite{he2020optimizing} & Delay and energy modelling & CNN \\ 

    \hline
    \multirow{18}{*}{Client} & Application to support users in everyday grocery shopping \cite{waltner2015mango} & Object classification & Random forests \\
    
    & Outdoor AR application for geovisualisation \cite{rao2017mobile}  & Object detection & SSD \\
    
    & MAR object detection \cite{li2020object}& Object detection & DNN \\
    
    & Reducing energy drain for MAR \cite{apicharttrisorn2019frugal} & Object detection & DNN \\ 
    
    & HoloLens surgical navigation \cite{von2021holoyolo}  & Object detection and pose estimation & CNN \\ 
    
    & AR inspection framework for industry \cite{perla2016inspectar} & Object detection & R-CNN \\
    
    & Learning Egyptian hierogplyphs with AR \cite{plecher2020arsinoe} & Object detection & SSD MobileNets \\

    & Interest point detection \cite{su2019deep} & Object detection & CNN \\
    
    & Edge-assisted distributed DNN for mobile WebAR \cite{ren2020edge} & Object recognition & DNN \\

    & AR navigation for landmark-based navigation \cite{amirian2016landmark} & Prediction of speed of movement & Penalised linear regression \\
    
    & Enhancing STEM education with AR \cite{ang2019enhancing} & Object detection & MobileNets \\
    
    & Campus navigation with AR \cite{lin2018novel} & 3D model placement determination & CNN \\
    
    & AR assisted process guidance on HoloLens \cite{redvzepagic2020sense} & Predicting process quality metrics & Decision tree classification \\
    
    & Indoor AR for Industry 4.0 smart factories \cite{subakti2018indoor} & Object detection & MobileNets \\
    
    & AR application for science education of nervous systems \cite{hoyle2019nervo} & Image classification & CNN \\ 
    
\hline
\end{tabular}
\end{table}

\subsection{ML for Object Detection and Object Recognition} \label{ssec:ml_objdet}
Several MAR systems include object detection and/or recognition as the primary pipeline task which uses ML methods during execution. Object detection supports MAR experiences by allowing the system to understand and detect which objects are in a particular scene at a given moment, and then return the spatial location and extent of each object instance \cite{liu2020deep}. For MAR in particular, object detection is used to provide environmental context-awareness and system knowledge of what supplemental augmentations should be provided users and where to place them on UIs. 

Deep learning are used to achieve object detection, where feature representations are automatically learned from data. The most commonly used type of deep learning models for object detection and object recognition are Convolutional Neural Networks (CNN) \cite{albawi2017understanding}, Region-based Convolutional Neural Networks (R-CNN) \cite{girshick2015region}, Faster Region-based Convolutional Neural Networks (Faster R-CNN) \cite{zhang2018terahertz}, You Only Look Once (YOLO) \cite{redmon2018yolov3}, Single Shot MultiBox Detectors (SSD) \cite{liu2016ssd}, ResNet \cite{he2015deep}, Neural Architecture Search Net (NASNet) \cite{zoph2016neural}, Mask Region-based Convolutional Neural Networks (Mask R-CNN) \cite{he2017mask}, DenseNet \cite{huang2017densely}, RetinaNet \cite{lin2017focal}, and EfficientNet \cite{tan2019efficientnet}. These networks can be used to create algorithms which are able to produce the same detection or recognition results. The major differences between each neural network are related to the structure and approach of the network itself when processing input data (e.g., camera images for optical object detection/recognition). For example, R-CNN, which tries to find rectangular regions which could contain objects in an image, based on the features of each rectangular region extracted by CNN, is one of the most important approaches for objective detection. Fast R-CNN has been proposed by applying the CNN to the whole image to find a rectangular region with fully connected layers to reduce computation cost caused by R-CNN. Fast R-CNN is proposed by merging R-CNN with region proposal network (RPN) to share full image convolutional features. Mask R-CNN is usually used in segmentation and could be regarded as a mixture of Faster R-CNN and fully convolutional networks for object detection. CNNs and the subsequent variants are therefore powerful components in the AR pipeline for producing AR experiences, without which, the system would require more time to analyse the environment which comes at the cost of user QoE. 

Both object detection and object recognition tasks in MAR systems are typically deployed on servers, namely, due to the limitation of client devices and their lack of computation and battery resources to sufficiently sustain acceptable MAR experiences. Examples of server-based deployment include usage of CNNs for object detection to support AR tracking \cite{akgul2016applying}, Mask R-CNN for object detection and instance segmentation to support smart task assistance for HoloLens-deployed AR \cite{park2020deep}, and CNN for object recognition in improving retail AR shopping experiences \cite{cruz2019augmented}. Additionally, some works explicitly state their usage of edge servers, citing the processing acceleration gained by the object detection and recognition tasks when a GPU is used to execute the functions, for example, using YOLO to accomplish object detection to support vehicle-to-edge AR \cite{zhou2019enhanced}, and SSD for supporting an edge-based AR platform for operators in production environments \cite{um2018modular}. Comparatively, there are several research works which deploy object detection and recognition in an enclosed client system, i.e., without needing to offload computation components to an external server or device. Amongst these works, both CNNs and DNNs are the most commonly attributed ML models. DNN models have been used to fulfil MAR object detection \cite{li2020object} and support the reduction in energy drain for MAR applications \cite{apicharttrisorn2019frugal}. Alternatively, CNNs are used for object detection and pose estimation for HoloLens surgical navigation \cite{von2021holoyolo}, interest point detection for an AR application \cite{su2019deep}, as well as for object detection in an industrial AR inspection framework \cite{perla2016inspectar}, albeit the model for this latter work is R-CNN. The deployment of these models on the client is supported by increasing efforts to produce more efficient neural networks, as well as reducing their implementation size through using ML libraries which are specifically designed for deployment on mobile devices, i.e., TensorFlow Lite \cite{tensorflow:lite}.

\subsection{ML for Object Tracking}
Object tracking in the MAR pipeline allows for objects in the physical world to be located and tracked across the MAR system, which subsequently supports the placement and tracking of virtual annotations derived from the analysis of the surrounding user environments in object detection and object recognition routines. As discussed in Section \ref{sec:framework}, two common used object tracking techniques are marker-based tracking and markerless tracking. Similar to object detection and object recognition, object tracking can be achieved through analysis of optical camera data \cite{weng2013objects}. However, object tracking differs from object detection as objects must be tracked over time while the MAR experience is occurring, whereas object detection is more concerned about detecting objects in every frame or when they first appear \cite{yilmaz2006object}.  

Some ML-based approaches for object tracking are Deep Regression Networks \cite{held2016learning}, Recurrent YOLO (ROLO) \cite{ning2017spatially}, Multi-Domain Networks (MDNet) \cite{nam2016learning}, Deep SORT \cite{wojke2017simple}, SiamMask \cite{wang2019fast}, TrackR-CNN \cite{voigtlaender2019mots}, Deep Regression Learning \cite{gao2018p2t}, Adversial Learning \cite{song2018vital}, Tracktor++ \cite{bergmann2019tracking}, and Joint Detection and Embedding (JDE) \cite{wang2019towards}. As evident by some of their names, these object tracking methods are based on deep learning models which are used for object detection and object recognition. For example, ROLO combines the object detection utility of a YOLO network with a Long Short-Term Memory (LSTM) network for acquiring the trajectory of an object. The combined ROLO network is able to track in both spatial and temporal domains, while dealing with occlusion and motion blur effects. TrackR-CNN uses Mask R-CNN for initial object detection, and then applies 3D convolutions to incorporate temporal context from the input data streams to form the object tracker, as well as detection and tracking. Meanwhile, TrackR-CNN is also able to perform segmentation within the single convolutional network. These techniques are not typically found in works which contain object tracking for an MAR system. Rather, custom computer vision tracking algorithms are combined with deep learning-based object detection and recognition models to track objects across MAR scenes. 

One such AR system which uses deep learning detection and a non-deep learning tracker is proposed by Ahmadyan \textit{et al.} \cite{ahmadyan2020instant}, where they use a CNN to initialise the pose of an object and then a planar tracker using sensor data to track the object in 9 degrees-of-freedom (DoF). The 3D tracking is performed on the client device, but utilises the mobile GPU to enhance the tracking performance and increase the rate the tracker is able to run at. Alternatively, a computer vision algorithm which is often used in MAR to map and track environments is SLAM, where sensor data from mobile devices such as gyroscopes and accelerometers are used to achieve reliable tracking \cite{piao2017adaptive}. Sernani \textit{et al.} \cite{sernani2019combining} use SLAM for a self-contained tourism AR application, where SLAM localisation allows the system to locate the user's orientation and assists in the superposition of arrows and icons to guide user's attention in an environment. On the other hand, Rambach \textit{et al.} \cite{rambach20186dof} investigate the usage of model-based tracking as an alternative to SLAM tracking as extracting real-world scale information and displaying object specific AR content may not always be possible with SLAM. Model-based tracking is achieved by using a predefined model or 3D reconstructed textured model of the tracking object, and then matching with a live view of the object during tracking to uncover the 6DOF pose of the object. Similar to previously mentioned MAR systems, they deploys the computation heavy object recognition and tracking tasks to a high performance edge server which is enabled with a GPU to reduce the total system latency, and the client is then free to solely perform data capturing and annotation rendering tasks.

Compared with purely deep learning-based object trackers, these latter systems fuse together a deep learning object detection routine with a non-deep learning tracking system. The usage of ML-based methods for tracking is less common when compared to object detection and recognition as there is generally a lack of tracking training data suitable for the creation of deep learning tracking models \cite{krebs2017survey}. However, exploratory works are now using recurrent neural network (RNN) for implicit learning of temporal dependencies during tracking from data.

\subsection{ML for Adaptive UIs}
In this section, we describe the ML methods applied to adaptive MAR UIs.

\textbf{InfoF}: 
The limitations of traditional information filtering methods are twofold \cite{MarkusTatzgern2016}, 
\begin{enumerate*}  \item the methods can be destructive and cause data loss for data visualisation, and \item the methods may fail when the amount of data grows tremendously and overloads.\end{enumerate*}
To avoid these limitations, using ML methods in layout generation and data clustering is the recent trend. The advantages of ML methods are summarised by Sahu \textit{et al.}~\cite{sahu2020artificial}, which is more efficient and effective rendering with ML-based layout creation, and secondly, better performance of transforming data into clustering-friendly nonlinear representations with deep learning-based data clustering. The information filtering, clustering, and rendering stages in MAR could be integrated as a context-aware recommendation system, with the ML methods for generating relevant recommendations for users based on the contextual situation of them. Recent works apply deep learning architectures for collaborative filtering in recommendation systems~\cite{bobadilla2020deep}. Their results show strong improvements in both prediction and recommendation of information for users as a result of incorporating context-awareness, user customisation, expert knowledge, and flexibility into recommendation systems~\cite{ebesu2017neural}. Context-awareness modules in AR can utilise the 5W1H context model to determine what kind of contextual information to be adopted~\cite{hong2007linking}. 
Adopting necessary context information within ML models and subsequently deciding which models to deploy, could reflect the logical thinking of the user, this is then a critical issue in AR recommendation systems.  
Jacucci \textit{et al.} \cite{jacucci2018combining} present a recommendation system for a mixed reality-based urban exploration application. They consider using user profiles, and weightings from both crowd-sourced ratings and personal preferences. Their ML technique for personalised search and recommendations is based on two main components, 
\begin{enumerate*} 
    \item a data model that defines the representation of multiple data sources (e.g., content, social, and personal) as a set of overlaid graphs, 
    and \item a relevance-estimation model that performs random walks with restarts on the graph overlays and computes a relevance score for information items. 
\end{enumerate*}
However, their study does not present any experimental results or performance data. Zhang \textit{et al.} \cite{zhang2013improving} present an aggregated random walk algorithm incorporating personal preferences, location information, and temporal information in a layered graph to improve recommendations in MAR ecosystems. Readme~\cite{chatzopoulos2016readme} is a real-time recommendation system for MAR with an online algorithm and by comparing this system with the following baselines: (i) k-Nearest Neighbours (kNN) which is based purely on physical distance, and (ii) focus-based recommendation which infers user’s interests based on their viewpoint and attempts to find the most similar objects within the current focus. The evaluation results show that ReadMe outperforms these baselines. Alternatively, contextual sequence modelling with RNN is another method to produce recommendation systems in MAR \cite{smirnova2017contextual}.

\textbf{RegEA}: 
Recent works show that applying ML algorithms in safety-critical AR systems could improve the tracking accuracy and lead to minimisations in the registration error. To adapt to real-time registration errors, safety-critical AR systems should provide some hints on the user interface to inform users of any potential harmful or dangerous situations. U-Net~\cite{ronneberger2015u} is a modified full CNN initially built for precise medical image segmentation. Brunet \textit{et al.} \cite{brunet2019physics} propose a physics-based DNN with U-Net, for providing AR visualisation during hepatic surgery. Physics-based DNN solves the deformed state of the organ by using only sparse partial surface displacement data and achieves a similar accuracy as a finite element method solution. They achieve the registration in only 3 ms with a mean target registration error (TRE) of 2.9 mm in the ex vivo liver AR surgery. Their physics-based DNN is 500× faster than a reference finite element method solution. For the RegEA visualisation, this work obtains the surface data of the liver with an RGB-D camera and ground truth data acquired at different stages of deformation using a CT scan. Markers were embedded in the liver to compute TRE. By knowing the TRE values, this work can highlight the region around the 3D model of the virtual liver to alert the user for the error minimisation. 

\textbf{Occlusion Representation and Depth Cues}:
Traditional methods for calculating the depth of a scene are ineffective in recovering both depth and scene information whenever perceptual cues are used. This is due to the heavy and under-constrained nature of the problem and the methods themselves, as well as the use of limited cues. The advantage of ML methods is their ability to exploit almost all cues simultaneously, thus offering better depth inference, as well as the capability to estimate depth information from RGBE images in a similar concurrent manner for the completion of other tasks, such as, object detection, tracking, and pose estimation~\cite{sahu2020artificial}. Park \textit{et al.}~\cite{park2020deep} present a comprehensive smart task assistance system for wearable AR, which adopts Mask R-CNN for several adaptive UI techniques. Their work supports occlusion representation calculation, registration error adaption, and adaptive content placement. Alternatively, Tang \textit{et al.} \cite{tang2020grabar} present GrabAR, a system using a custom compact DNN for generating occlusion masks to support real-time grabbing of virtual objects in AR. The model is able to calculate and segment the hand to provide visually plausible interactions of objects in the virtual AR environment. 

\textbf{IEsti}: 
The major limitation of classical methods for illumination estimation is the low accuracy problem due to insufficient contextual information, especially the methods using auxiliary information, the methods are computationally expensive and not scalable, which disturbs real-time performance of AR scene rendering. ML-based illumination estimation resolves the low accuracy and performance issue in AR, which does not require definitive models of the geometric or photometric attributes~\cite{sahu2020artificial}. DNN approaches are based on the assumption that prior information about lighting can be learned from a large dataset of images with known light sources. Large datasets of panoramas are used to train an illumination predictor, they use the concept of finding similarities between an input image and one of the projections of individual panoramas~\cite{karsch2014automatic}. Gardner \textit{et al.}~\cite{gardner2017learning} present a illumination estimation solution by calculating diffuse lighting in the form of an omnidirectional image using a neural network. DeepLight~\cite{kan2019deeplight} is a CNN-based illumination estimation for calculating a dominant light direction from a single RGB-D image. DeepLight uses a CNN to encode a relation between the input image and a dominant light direction. DeepLight applies outlier removal and temporal filtering to achieve temporal coherence of light source estimation. The results of DeepLight are evaluated with a comparison to the Gardner \textit{et al.}~\cite{gardner2017learning} approach. DeepLight achieves a higher accuracy for the estimated light direction than the approach of Gardner \textit{et al.}~\cite{gardner2017learning} by calculating the angular error to the ground-truth light direction and by rendering virtual objects in AR using an estimated light source. LightNet~\cite{nathan2020lightnet} is the latest illumination estimation using a dense network (DenseNet) architecture, which trained with two softmax outputs for colour temperature and lighting direction prediction. 
\section{Research Challenges and Further Directions} \label{sec:future_research}
We discuss some prominent challenges for MAR, including seamless frameworks for interactable immersive environments, audio AR, user interactions, user interactions with environments, and federated learning. It is crucial to identify and analyse these challenges and seek for novel theoretical and technical solutions. 

\subsection{Towards seamless frameworks for interactable immersive environments}
Existing frameworks and SDKs offer various features to understand physical environments. \textit{ARKit} remarkably achieves a broad coverage of feature recognition, including point clouds, anchors, light estimation, and occlusion. In general, the AR content annotations in existing frameworks already go beyond simple overlays in mid-air. Feature recognition enables these augmented annotations to become a part of the physical world, albeit users can distinguish unnatural positioning and some rendering issues. Interactive AR on human faces, such as with Banuba and XZIMG, demonstrates a real-life use case that requires AR overlays to merge with the physical world. We see an obvious trend for the merging of AR augmentations in the physical world for different application scenarios, e.g., for medical surgery~\cite{brunet2019physics}. One potential research direction could be to extend the different sensor types and employ multi-sensor information to construct user context intelligently. 

Alternatively, ML approaches can serve as an automated tool to automatically generate lively and animated digital entities~\cite{holden2019-physics} as well as imitating real world creatures~\cite{Schulman2016HighDimensionalCC, Tassa2012Trajectory}. In addition to this, the collection of ML-generated objects can potentially become a library of animated objects for AR. However, \textit{``how to merge the digital overlaid content (i.e., AR content) interactions with the physical world (e.g., everyday objects and humans)''} are in the nascent stage. The merged AR overlays will eventually co-exist with both human users and their surroundings. This leaves us with an unexplored research challenge of designing novel spatial environments. Sample research directions for the next generation of MAR frameworks could be: 
\begin{enumerate*}
    \item user issues such as immersive spatial environments and their relationship and interaction among human users, virtual entities, and tangible objects;
    \item AI becomes a key actor in spatial environments and how objects (or agents) take actions in an automated fashion;
    and \item how to evaluate the interaction of AI-supported animated objects with physical surroundings (i.e., human users and physical objects) without active participation of designers and software engineers (analogous to the concept of \textit{evaluation probe}~\cite{Timo2015-Probe}). 
\end{enumerate*}

\subsection{Audio augmented reality}
While the majority of existing frameworks have made significant efforts in visual-based AR cues, another neglected and important aspect is audio-based interaction. Audio can be regarded as an ineluctable design material for audio-based interactions, where individualisation, context-awareness, and diversification are the primary strategies of designing such audio-based interaction~\cite{sutton2019}. Among three strategies, context awareness (i.e., voices to be tailored to the user's context) aligns with the primary aspect of AR. In addition, audio's sociophonetic aspect should reflect the user's social quality. For example, the scenario of paying a credit card bill requires audio that brings the user an impression of trustworthiness, honesty, and efficiency~\cite{sutton2019}. Audio AR will be an interesting topic in examining the alignment of audio and the properties of virtual visuals (e.g., appearance and other contexts) in such audio-driven AR. 

\subsection{User interaction and user workload}
Section~\ref{ssec:framework_features} summarises 37 MAR frameworks and SDKs. Generally, these frameworks have not prioritised natural user interfaces (NUIs)~\cite{lee2018interaction} and very limited numbers of existing frameworks and SDKs support on-body interfaces such as hand tracking, facial tracking, and body tracking. Although such feature tracking on user bodies enables alternative input modals on AR devices, especially head-worn AR headsets, such discussed frameworks and SDKs do not offer sufficient considerations into the user workload across various types of user interaction tasks, ranging from clicking a button on a 2D menu to manipulating a 3D sphere with rotation movements. More importantly, there exists no integration of user workload measurements into existing frameworks. 

As natural user interfaces rely on body movements, user interaction with such interfaces cannot be prolonged, due to user fatigue~\cite{lee2018interaction, lee2019hibey}. It is necessary to include computational approaches to understand how the user interacts with AR interfaces to reduce user workload (i.e., collecting user interaction traces). The user interaction trace will be collected for the foundation of data-driven user interaction design, which further offers systematic evaluation modules inside MAR frameworks and SDKs to optimise the user interaction design. The systematic evaluation modules should also acquire the ability to assist the iterative design process that allows for natural user interfaces to improve the user workloads with constantly changing AR environments.

\subsection{Users and their environments}
As mentioned above, with user interaction and the potential architecture of collecting data user traces, as these facilities are becoming mature, such frameworks can further consider the user interaction with their tangible environments. This then supports hyper-personalisation of user experiences, i.e., \textit{letting audiences shape their user interaction designs}. 
Users with AR devices, e.g., smartphones (as-is) and smartglasses (to-be), encounter digital entities that are highly customised to the user preference. It is worthwhile mentioning that digital entities will merge with the nearby tangible environments and user interaction in AR should work on both physical and digital worlds. 

With such premises, the next generation of AR frameworks will encounter two primary challenges. On the one hand, the tangible environments are not static objects. The information and property registration of such objects will not remain static forever, and any changes in such objects will largely impact the user experience. On the other hand, we cannot guarantee that users will act in the same identical way when interacting with AR augmentations on tangible environments, as initially planned by software engineers and product managers when they designed and implemented the AR physical environments, which presumably considered the world to be a static property. 

The second challenge is more challenging to be addressed compared to the first one, because the first will eventually be solved by the enhanced sensing capability of our tangible environments and the crowd-based mechanisms that can constantly monitor and reflect the changes in the tangible environments. Regarding the second challenge, we cannot blame such user misalignment between the real user behaviours and the intended design as human error.
We have the following example, derived from a classic design example of ``\textit{the desired path}'', to reinforce the above statement. In a school campus, a user is finding a way to reach a certain building, and there exists two official paths (left or right) and a lawn in front of the user. The user with an AR device receives navigational information of either `going onto the path on the left-hand side' or `going along the path on the right-hand side'. Such navigation information is contextualised with the school campus map and registered in the design and implementation phases. However, due to the incentive of the shortest path, the user eventually takes the diagonal (unofficial) path across the lawn, instead of taking any of the official paths. Meanwhile, the AR system keeps reminding the user about the erroneous path being taken, according to such pre-defined AR information (referring to the latter challenge). In addition, if such a diagonal path has been repetitively taken, the tangible environment has changed, but the updates of AR information lags behind (referring to the first challenge).

Such uncertainties from dynamic tangible environments and user behaviours give the motivation to building adaptive information management in AR frameworks. Although we have no definitive solutions to address the above challenges at the time being, we identify a research gap in fulfilling the orientation towards user interaction and their environments in adaptive and personalised manners.
 
\subsection{Deepening the understanding of users' situations} 
Nowadays, the frameworks and SDKs provide well-established connections to common sensors such as cameras, inertial measurement units (IMUs), and GPS sensors. The sensing capability allows AR frameworks or systems to detect objects and people within the user's in-situ environments, regardless of marker-based (Fig.~\ref{fig:tracking_examples}) or markerless approaches. Accordingly, the AR framework makes timely responses and offers user feedback (visual, audio, and haptics), e.g., 3D images popping up on the user's screen. Although research prototypes attempt to manage the user feedback by leveraging the user's geographical information~\cite{woo2011development}, many existing applications are limited by a moderate level of tracking precision due to GPS. 

When AR frameworks are being deployed on a large scale, we foresee the number of candidate AR feedback in one specific location (no less than $5\times5$ m$^2$) will surge exponentially. In such a way, AR frameworks may encounter a bottleneck in delivering optimised information flow (i.e., concise and context-aware visuals) on AR devices. In other words, such geographical location-based tracking requires auxiliary channels (e.g., camera-based detection) to get a more precise and detailed understanding of user's environments that guarantee the performance of adaptive user interfaces~\cite{lam2019m2a}. In return for improved detection and tracking precision, the framework will take additional overheads. However, the camera-based detection and tracking generate offloading tasks that demand high network traffic, and hence produces significant latency from the distant cloud-based servers. As AR users will be impacted significantly by degraded QoE, we have to restructure the client-server architectures in AR frameworks and SDKs.

\subsection{Privacy preserving MAR frameworks}
Multi-user collaboration has been rarely considered by existing frameworks. One key reason is that such remote collaborations may introduce privacy risks. Sharing user information in effective and just-in-need manners will be a grand challenge. We summarise several challenges to address potential privacy leakage in multi-user AR scenarios, as follows. First, the next generation MAR frameworks will need a content-sharing module to expose any sensitive information among multiple users across various applications. One remarkable example is the sharing of user locations in physical space, in which multiple users work on the same AR spatial environment. Some users may own their preferred privacy policies and hence will threaten such location-based user collaborations. Second, privacy policies from multiple users can conflict with each other. In other words, a user does not realise that one's action could deteriorate the privacy policy of another user in AR, primarily subject to the spatio-temporal relationship. Third, user interaction with AR objects could pose privacy leakage due to its historical record, namely user interaction trace, especially when we consider AR objects as some public properties in this emerging spatial environments. For example, a user may establish some virtual stickers and share to target users in an AR spatial environment. However, different users may own asynchronous views and hence unparalleled information due to their privacy preference, and hence they work differently on the virtual stickers due to different beliefs. Thus, the effectiveness of preventing privacy leakage highly depends on not only the user-centric and user-interdependent semantics on the application level, bu also a multi-user framework level. It is important to strengthen the information-theoretic dimension in AR frameworks.

\subsection{Federated learning for MAR}
The traditional cloud-based MAR frameworks distribute the centralised trained models to AR clients, which may require prohibitively large amounts of time when models are often updated. Federated learning is an emerging ML paradigm to train local models in a decentralised manner and can cooperate with any ML method, which supports personalised models for users and significantly reduces the communication latency. Instead of centralising the data to train a global model for improved local predictions, federated learning downloads the global model to capable devices, where models are locally updated. Updates are sent to a central server, which then aggregates the local updates into an updated global model~\cite{chen2020federated}.

For example, to address rigorous network constraints of MAR systems, most existing MAR solutions focus on viewport adaptive schemes, which segments a video into multiple tiles. The system assesses the client's viewport position in the video and predicts the most probable future viewport positions. The server then only transmits the tiles located in the field of view of the user and the tiles at the predicted viewport locations. A federated learning approach can train a local viewport prediction model and only transmits the model updates to the server, and such approaches that perform viewport prediction on-device and requests the corresponding tiles from the server. Therefore, federated learning-based approach allows training the global model in a distributed fashion while enabling fine-grained personalisation through the personal model.
\section{Conclusions} \label{sec:conclusions}
This survey employs a top-down approach to review MAR frameworks supporting user interfaces and applications. We begin with the discussion of MAR applications, and the content management and generation that support AR visualisation adaptive to the user's environments of high dynamics and mobility. The article then meticulously selects 37 MAR frameworks either available in industry or proposed by researchers in academia, and further reviews and compares their functions and features. Additionally, the article covers the MML methods in the domain of MAR and provides a trendy analysis of how the methods support MAR becoming sensitive to user contexts and physical environments. We discussed the major challenges and unexplored topics of designing seamless yet user-centric MAR frameworks, potentially empowered by ML methods. Finally, this article hopes to provide a broader discussion within the community, and intends to invite researchers and practitioners, especially in the fields relevant to MAR, to shape the future of AR. 
 
\bibliographystyle{ACM-Reference-Format}
\bibliography{bibliographies/bibliography_reduced, bibliographies/jamesbibliography_reduced}

\end{document}